\documentclass[11pt,a4paper]{article}

\usepackage{jheppub}
\usepackage{amsmath}
\usepackage{amssymb}
\usepackage{graphics}
\usepackage[active]{srcltx}
\usepackage{pdfsync}
\usepackage{shuffle}
\usepackage{slashed}
\usepackage{hyperref}
\usepackage{subfigure}

\newcommand{\bea}{\begin{eqnarray}}
\newcommand{\eea}{\end{eqnarray}}
\newcommand{\nn}{\nonumber \\}

\def\W #1{\widetilde{#1}}

\def\eref#1{(\ref{#1})}

\def\a{{\alpha}}

\def\b{{\beta}}

\allowdisplaybreaks

\title{$2$-split from Feynman diagrams and Expansions}
\author[a,b,c,d]{Bo Feng}\emailAdd{fengbo@scnu.edu.cn}
\author[c]{Liang Zhang}\emailAdd{liangzhang@csrc.ac.cn}
\author[e]{Kang Zhou}\emailAdd{zhoukang@yzu.edu.cn}

\affiliation[a]{State Key Laboratory of Nuclear Physics and Technology, Institute of Quantum Matter, South China Normal University, Guangzhou 510006, China}
\affiliation[b]{Guangdong Basic Research Center of Excellence for Structure and Fundamental Interactions of Matter, Guangdong Provincial Key Laboratory of Nuclear Science, Guangzhou 510006, China}
\affiliation[c]{Beijing Computational Science Research Center, Beijing 100084, China}
\affiliation[d]{Peng Huanwu Center for Fundamental Theory, Hefei, Anhui, 230026, China}
\affiliation[e]{Center for Gravitation and Cosmology, College of Physical Science and Technology, Yangzhou University, No.180, Siwangting Road, Yangzhou, 225009, China}

\date{\today}
\abstract{In this paper, we investigate the $2$-split behavior of tree-level amplitudes of bi-adjoint scalar (BAS), Yang-Mills (YM), non-linear sigma model (NLSM), and general relativity (GR) theories under certain kinematic conditions. Our approach begins with a proof, based on the Feynman diagram method, of the $2$-split property for tree-level BAS$\oplus$X amplitudes with $\mathrm{X}={\mathrm{YM},\mathrm{NLSM},\mathrm{GR}}$. The proof relies crucially on a particular pattern in the Feynmam rules of various vertices. Building on this, we use the expansion of X amplitudes into  BAS$\oplus$X amplitudes to establish
the $2$-split behavior.  As a byproduct, we derive universal expansions of the resulting pure X currents into BAS currents, which closely parallel the corresponding on-shell amplitude expansions.
 }

\begin{document}

\maketitle \flushbottom

\section{Introduction}
\label{sec-intro}

In the past two years, there has been an avalanche of progress in the study of non-supersymmetric scattering amplitudes, extending geometric and combinatorial ideas to theories that more closely resemble the real world \cite{Arkani-Hamed:2023lbd,Arkani-Hamed:2023mvg,Arkani-Hamed:2023swr,Arkani-Hamed:2023jry,Arkani-Hamed:2024nhp,Arkani-Hamed:2024vna,Arkani-Hamed:2024yvu,Arkani-Hamed:2024tzl,Arkani-Hamed:2024nzc,Arkani-Hamed:2024pzc}. Motivated by the discovery of factorization near zero in these works, a novel tree-level phenomenon—dubbed the $2$-split—was uncovered in \cite{Cao:2024gln,Cao:2024qpp} and subsequently explored from various perspectives \cite{Arkani-Hamed:2024fyd,Zhou:2024ddy,Feng:2025ofq}. The $2$-split describes the factorization of certain tree-level amplitudes into two amputated currents on special loci in kinematic space, and it exhibits remarkable universality across a wide range of field-theoretic and string-theoretic models. A related but distinct factorization behavior in YM amplitudes was proposed in \cite{Zhang:2024iun,Zhang:2024efe}. Meanwhile, the closely connected phenomenon of hidden zero—the vanishing of amplitudes on similar loci—has been extensively investigated \cite{Arkani-Hamed:2023swr,Rodina:2024yfc,Bartsch:2024amu,Li:2024qfp,Zhang:2024efe,Huang:2025blb,Feng:2025ofq}.

As with any newly uncovered universal behavior, a crucial next step is to probe it from multiple angles to develop a deeper conceptual understanding. To date, four complementary perspectives on the $2$-split have been proposed:
\begin{enumerate}
	\item    Extracting the $2$-split from the factorization of measures and integrands in stringy formulas \cite{Cao:2024gln,Cao:2024qpp};
	\item  Viewing it as cutting a large surface into two smaller ones \cite{Arkani-Hamed:2024fyd};
	 \item  Reducing it via BCFW recursion to the trivial $2$-split of the lowest-point amplitudes \cite{Feng:2025ofq};
	\item  Interpreting it as splitting propagators along special lines in each Feynman diagram for ${\rm Tr}(\phi^3)$ theory \cite{Zhou:2024ddy}.
\end{enumerate}

Each of the four approaches not only offers new conceptual and pictorial insights but also opens avenues for extending the $2$-split to tree amplitudes in a broader class of theories, as well as to loop-level Feynman integrands. In this paper, we focus on the fourth approach and further develop for other theories, such as YM, NLSM and GR.

The diagrammatic construction of \cite{Zhou:2024ddy} rests on a simple but universal observation: in any Feynman diagram, starting from three external legs $i$, $j$, and $k$, one can always identify three lines $L_{i,v}$, $L_{j,v}$, and $L_{k,v}$ meeting at a common vertex $v$. This structure exists for all diagrams under consideration. The $2$-split can then be realized by splitting the propagators along $L_{i,v}$ and $L_{j,v}$ in each diagram. While this approach works efficiently for pure ${\rm Tr}(\phi^3)$ amplitudes, its direct application to YM amplitudes becomes significantly more intricate, with the complexity growing rapidly as the number of distinct vertex types increases. As expected, in theories with infinitely many vertices—such as the NLSM and GR—a purely Feynman-rule-based implementation is not practical.

Nevertheless, as we will show, the method can be naturally generalized to tree-level BAS amplitudes—more general than ${\rm Tr}(\phi^3)$ amplitudes—as well as to certain mixed BAS$\oplus$X amplitudes with a sufficiently large number of external BAS scalars, where $\mathrm{X}={\mathrm{YM},\mathrm{NLSM},\mathrm{GR}}$. This suggests a broader strategy: if complicated pure X amplitudes can be expanded into such BAS$\oplus$X amplitudes, then their $2$-split may follow directly from the $2$-split of the BAS$\oplus$X building blocks.

Such decompositions can be achieved via universal amplitude expansions, extensively studied in \cite{Stieberger:2016lng,Schlotterer:2016cxa,Chiodaroli:2017ngp,Nandan:2016pya,delaCruz:2016gnm,Fu:2017uzt,Teng:2017tbo,Du:2017kpo,Du:2017gnh,Feng:2019tvb,Zhou:2019gtk,Zhou:2019mbe,Wei:2023yfy,Hu:2023lso,Du:2024dwm,Zhou:2024qwm,Zhou:2024qjh}, which express tree amplitudes in a wide variety of theories in terms of BAS amplitudes with universally defined coefficients. These expansions have been fruitfully applied to the study of the hidden zero phenomenon \cite{Huang:2025blb}, where hidden zeros of diverse theories are reduced to those of BAS amplitudes—easily verified using the diagrammatic method of \cite{Zhou:2024ddy}. However, the specific Kleiss–Kuijf–based techniques employed in \cite{Huang:2025blb} do not carry over to $2$-split behavior. In contrast, under $2$-split constraints, universal expansions naturally yield decompositions into BAS$\oplus$X amplitudes, whose $2$-split can again be derived diagrammatically.

In this work, first we use the diagrammatic method to establish $2$-split for tree-level BAS$\oplus$X with $\mathrm{X}={\mathrm{YM},\mathrm{NLSM},\mathrm{GR}}$. Then we use the universal expansions of pure X theory
to establish $2$-split for tree-level pure X theory. In these factorizations, each X amplitude decomposes into a pure X current and a mixed BAS$\oplus$X current with three external BAS scalars (equivalently, a ${\rm Tr}(\phi^3)\oplus$X current with three $\phi$’s). As a byproduct, universal expansions of the resulting pure X currents into BAS currents have also be derived, which closely mimic the corresponding on-shell amplitude expansions.


The remainder of this paper is organized as follows. In Section~\ref{sec-BAS}, we review the diagrammatic method of \cite{Zhou:2024ddy} and extend it to BAS amplitudes. Section~\ref{sec-BAS+X} further generalizes this method to obtain $2$-split of special BAS$\oplus$X amplitudes. A concise review of universal expansions is given in Section~\ref{sec-expansion}. In Section~\ref{sec-YM-NLSM}, we combine the $2$-split derived in Section~\ref{sec-BAS+X} with universal expansions in Section~\ref{sec-expansion} to obtain $2$-split of pure X amplitudes. Finally, Section~\ref{sec-summary} summarizes our results and discusses possible extensions.

\section{$2$-split of tree-level ${\rm Tr}(\phi^3)$ and BAS amplitudes}
\label{sec-BAS}

The simplest manifestation of the $2$-split behavior occurs in ${\rm Tr}(\phi^3)$ amplitudes. As demonstrated in \cite{Zhou:2024ddy}, this behavior can be established diagrammatically using only the Feynman rules. In this section, we extend the method of \cite{Zhou:2024ddy} to the more general cubic scalar theory known as the BAS theory, and demonstrate the $2$-split for doubly ordered partial BAS amplitudes.

The BAS theory describes cubic interactions of massless scalars, governed by the Lagrangian
\bea
{\cal L}_{\rm BAS} = \frac{1}{2}\partial_\mu\phi^{Aa}\partial^{\mu}\phi^{Aa} + \frac{\lambda}{3!}F^{ABC}f^{abc}
\phi^{Aa}\phi^{Bb}\phi^{Cc}\, \label{Lag-BAS}
\eea
where $F^{ABC} = {\rm tr}([T^A, T^B] T^C)$ and $f^{abc} = {\rm tr}([T^a, T^b] T^c)$ are the structure constants of two distinct Lie groups. Each scalar field $\phi^{Aa}$ carries a pair of group indices.

By employing standard techniques for group factor decomposition, the tree-level BAS amplitude can be expressed as
\bea
A^{\rm BAS}_n=\sum_{\sigma\in{\cal S}_n\setminus Z_n}\,\sum_{\sigma'\in{\cal S}'_n\setminus Z'_n}\,
{\rm tr}[T^{A_{\sigma_1}},\cdots T^{A_{\sigma_n}}]\,{\rm tr}[T^{a_{\sigma'_1}}\cdots T^{a_{\sigma'_n}}]\,
{\cal A}^{\rm BAS}_n(\sigma_1,\cdots,\sigma_n|\sigma'_1,\cdots,\sigma'_n)\,,\label{2.2}
\eea
where $A_n$ denotes the full $n$-point tree-level amplitude with coupling constants factored out. The sums run over all cyclically inequivalent permutations of external scalar legs, denoted by ${\cal S}_n \setminus Z_n$ and ${\cal S}'_n \setminus Z'_n$. Each partial amplitude ${\cal A}^{\rm BAS}_n(\sigma_1, \ldots, \sigma_n \,|\, \sigma'_1, \ldots, \sigma'_n)$ includes only massless scalar propagators and is planar with respect to both orderings.

For instance, the four-point partial amplitude ${\cal A}^{\rm BAS}_4(1,2,3,4\,|\,1,2,4,3)$ is given by
\bea
{\cal A}^{\rm BAS}_4(1,2,3,4|1,2,4,3)={1\over s_{12}}\,,
\eea
up to an overall sign. Here and throughout, the Mandelstam variable is defined as
\bea
s_{p\cdots q}\equiv k_{p\cdots q}^2\,,~~~{\rm with}~k_{p\cdots q}\equiv\sum_{i=p}^q\,k_i\,,~~~~\label{mandelstam}
\eea
In this example, the amplitude contains only the $1/s_{12}$ channel, as the Feynman diagrams corresponding to $1/s_{14}$ and $1/s_{13}$ are incompatible with both orderings $(1,2,3,4)$ and $(1,2,4,3)$.

The antisymmetry of the structure constants $F^{ABC}$ and $f^{abc}$ implies that each partial amplitude carries an overall sign, which arises from the reordering of external legs on the vertices. This sign can be determined by counting the number of flips between the two orderings, as discussed in \cite{Cachazo:2013iea}. For notational convenience, we occasionally represent an $n$-point partial amplitude as ${\cal A}^{\rm BAS}_n(\boldsymbol{\sigma}_n\,|\,\boldsymbol{\sigma}'_n)$, where $\boldsymbol{\sigma}_n$ and $\boldsymbol{\sigma}'_n$ denote the two orderings.

On special loci in kinematic space, certain partial amplitudes in the BAS theory exhibit a novel factorization property known as the $2$-split behavior. Concretely, consider an $n$-point BAS amplitude where three external legs $\{i, j, k\}$ are selected, and the remaining $n-3$ legs are partitioned into two disjoint subsets $A$ and $B$, such that $A \cup B = \{1, \ldots, n\} \setminus \{i, j, k\}$. If the two orderings $\boldsymbol{\sigma}_n$ and $\boldsymbol{\sigma}'_n$ of a partial amplitude ${\cal A}_{\rm BAS}(\boldsymbol{\sigma}_n\,|\,\boldsymbol{\sigma}'_n)$ satisfy the condition
\bea
{\pmb\sigma}_n=i,\{\pmb A,j,\pmb B\}\shuffle k\,,~~~~{\pmb\sigma}'_n=i,\{\pmb A',j,\pmb B'\}\shuffle' k\,,~~~~{\rm up~to~cyclic~permutations}\,,~~\label{compa-order}
\eea
then the amplitude factorizes as
\bea
& & {\cal A}^{\rm BAS}_n(i,\pmb A,j,\pmb B\shuffle k|i,\pmb A',j,\pmb B'\shuffle' k)\xrightarrow[]{\eref{kinematic-condi-split-phi3}}\nn
&&{\cal J}^{\rm BAS}_{n_1}(i,\pmb A,j,\kappa|i,\pmb A',j,\kappa)\,\times\,{\cal J}^{\rm BAS}_{n+3-n_1}(j,\pmb B\shuffle\kappa^*,i|j,\pmb B'\shuffle'\kappa^*,i)\,,~~\label{split-BAS}
\eea
or
\bea
& & {\cal A}^{\rm BAS}_n(i,\pmb A\shuffle k,j,\pmb B|i,\pmb A',j,\pmb B'\shuffle' k)\xrightarrow[]{\eref{kinematic-condi-split-phi3}}\nn
&&{\cal J}^{\rm BAS}_{n_1}(i,\pmb A\shuffle\kappa,j|i,\pmb A',j,\kappa)\,\times\,{\cal J}^{\rm BAS}_{n+3-n_1}(j,\pmb B,i,\kappa^*|j,\pmb B'\shuffle'\kappa^*,i)\,,~~\label{split-BAS-2}
\eea
depending on $\pmb\sigma_n$ and $\pmb\sigma'_n$, on the locus
\bea
s_{ab}=0\,,~~~~{\rm with}~a\in A\,,~b\in B\,.~~\label{kinematic-condi-split-phi3}
\eea

Here, $\boldsymbol{A}$, $\boldsymbol{A}'$, $\boldsymbol{B}$, and $\boldsymbol{B}'$ are ordered permutations of the sets $A$ and $B$, respectively. Throughout this discussion, bold symbols denote ordered sets. The shuffle notation $\shuffle$ denotes all interleavings of two ordered sets $\boldsymbol{\alpha}$ and $\boldsymbol{\beta}$ that preserve the internal order within each set. For example, if $\boldsymbol{\alpha} = \{1,2\}$ and $\boldsymbol{\beta} = \{3,4\}$, then $\boldsymbol{\alpha} \shuffle \boldsymbol{\beta}$ includes:
\bea
\{1,2,3,4\}\,,~~~~\{1,3,2,4\}\,,~~~~\{1,3,4,2\}\,,~~~~\{3,4,1,2\}\,,~~~~\{3,1,4,2\}\,,~~~~\{3,1,2,4\}\,.
\eea

As illustrative examples of \eqref{compa-order}, \eqref{split-BAS}, and \eqref{split-BAS-2}, suppose $\boldsymbol{A} = \{a\}$, $\boldsymbol{B} = \{b_1, b_2\}$. Then $\boldsymbol{\sigma}_6 = \{i, \boldsymbol{A}, j, \boldsymbol{B}\} \shuffle k$ includes orderings such as $(i, k, a, j, b_1, b_2)$, $(i, a, k, j, b_1, b_2)$, $(i, a, j, k, b_1, b_2)$, $(i, a, j, b_1, k, b_2)$, and $(i, a, j, b_1, b_2, k)$; while $\boldsymbol{\sigma}'_6 = i, \boldsymbol{A} \shuffle k, j, \boldsymbol{B}$ allows $(i, k, a, j, b_1, b_2)$ and $(i, a, k, j, b_1, b_2)$. For convenience, we refer to such orderings as compatible orderings with the kinematic constraint \eqref{kinematic-condi-split-phi3}, meaning that elements in $A$ and $B$ are separated by legs $i$ and $j$.

In the factorized current ${\cal J}^{\rm BAS}_{n_1}(i, \boldsymbol{A}, j, \kappa\,|\,i, \boldsymbol{A}', j, \kappa)$ in \eqref{split-BAS}, the leg $\kappa$ carries off-shell momentum $k_\kappa = k_k + k_B$, as required by momentum conservation. Correspondingly, in ${\cal J}^{\rm BAS}_{n+3-n_1}(j, \boldsymbol{B} \shuffle \kappa^*, i\,|\,j, \boldsymbol{B}' \shuffle' \kappa^*, i)$, the leg $\kappa^*$ carries momentum $k_{\kappa^*} = k_k + k_A$, and occupies the position originally held by leg $k$ in both orderings. A similar interpretation applies to the factorization in \eqref{split-BAS-2}.

To clarify the notations, we present two illustrative examples. In the first example, consider the 5-point case with the choice $i=1$, $j=3$, $k=5$, $A=\{2\}$, and $B=\{4\}$. The corresponding amplitude factorizes as

\bea
{\cal A}^{\rm BAS}_5(1,2,3,4,5|1,2,3,5,4)\xrightarrow[]{s_{25}=0}{\cal J}^{\rm BAS}_4(1,2,3,\kappa|1,2,3,\kappa)\,\times\,{\cal J}^{\rm BAS}_4(3,4,\kappa^*,1|3,\kappa^*,4,1)\,.
\eea

In the second example, we consider the 6-point case with $i=2$, $j=4$, $k=5$, $A=\{3\}$, and $B=\{1,6\}$. The amplitude then factorizes as

\bea
&& {\cal A}^{\rm BAS}_6(1,2,3,5,4,6|1,5,2,3,4,6)\xrightarrow[]{s_{13}=s_{36}=0}\nn
&&{\cal J}^{\rm BAS}_4(2,3,\kappa,4|2,3,4,\kappa)\,\times\,{\cal J}^{\rm BAS}_5(4,6,1,2,\kappa^*|4,6,1,\kappa^*,2)\,.
\eea

The $2$-split in \eref{split-BAS} and \eref{split-BAS-2} were established in \cite{Cao:2024qpp} using the CHY formalism. We now prove them by directly applying Feynman rules.

\begin{figure}[ht]
	\centering
	\includegraphics[width=16cm]{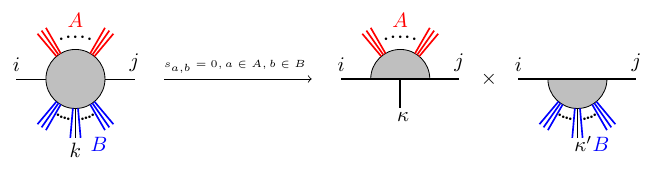} \\
	\caption{$2$-split of ${\rm Tr}(\phi^3)$ amplitude. }\label{split}
\end{figure}

\begin{figure}[ht]
	\centering
	\includegraphics[width=15cm]{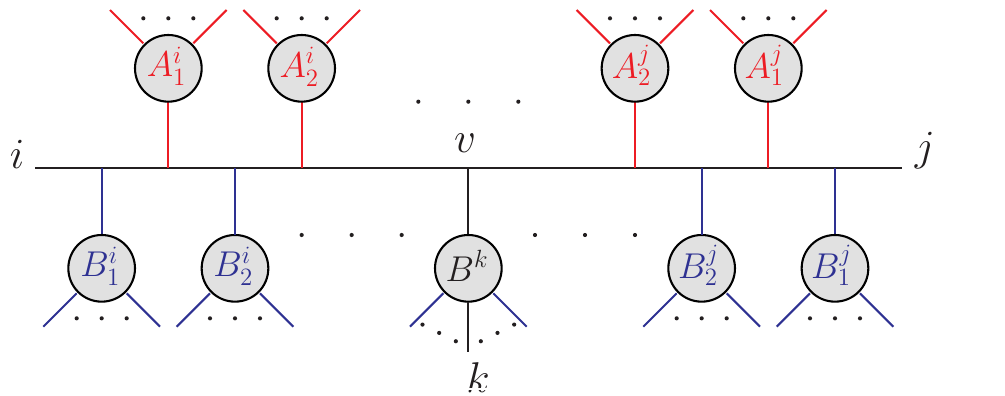} \\
	\caption{The common structure for each diagram. }\label{npphi3}
\end{figure}

To begin, we consider a special case where the two orderings $\boldsymbol{\sigma}_n$ and $\boldsymbol{\sigma}'_n$ are identical. The corresponding amplitudes are referred to as ${\rm Tr}(\phi^3)$ amplitudes. In this case, the two-split of an $n$-point ${\rm Tr}(\phi^3)$ amplitude takes the form

\bea
{\cal A}^{{\rm Tr}(\phi^3)}_n(i,\pmb A,j,\pmb B\shuffle k)&\xrightarrow[]{\eref{kinematic-condi-split-phi3}}&{\cal J}^{{\rm Tr}(\phi^3)}_{n_1}(i,\pmb A,j,\kappa)\,\times\,{\cal J}^{{\rm Tr}(\phi^3)}_{n+3-n_1}(j,\pmb B\shuffle\kappa^*,i)\,,~~\label{zero-phi3}
\eea
as depicted in Fig.~\ref{split}.
We now briefly review the diagrammatic proof presented in \cite{Zhou:2024ddy} for the aforementioned non-trivial factorization behavior. For any diagram contributing to  ${\cal A}^{{\rm Tr}(\phi^3)}_n(\boldsymbol{A},i,\boldsymbol{B} \shuffle k,j)$, one can always identify a central vertex, $v$, where the lines associated with external legs $i$, $j$, and $k$ (denoted $L_{i,v}$, $L_{j,v}$, and $L_{k,v}$). Each diagram can be viewed as attaching sub-blocks to $L_{i,v}$ and $L_{j,v}$, as illustrated in Fig.~\ref{npphi3}. 

In Fig.~\ref{npphi3}, each external leg $a \in \boldsymbol{A}$ belongs to some block $A_c^\ell$, with $c \in \{1,\dots,m^\ell\}$, and $\ell = i,j$. Similarly, each $b \in \boldsymbol{B}$ belongs to a block $B_d^\ell$ or $B^k$, where $d \in \{1,\dots,h^\ell\}$. The full amplitude can thus be expressed as
\bea
{\cal A}^{{\rm Tr}(\phi^3)}_n(1,\cdots,n)&=&\sum_{\rm division}\,{\cal J}_{B^k}\,\sum_{\Gamma}\,\Big[\prod_{\ell=i,j}\,\Big(\prod_{c=1}^{m_\ell}\,{\cal J}_{A^\ell_c}\Big)\,
\Big(\prod_{d=1}^{h_\ell}\,{\cal J}_{B^\ell_d}\Big)\,\Big(\prod_{t=1}^{m_\ell+h_\ell}{1\over{\cal D}^{\ell,v}_t}\Big)\,\Big]\,,~~\label{np-for0}
\eea
where ${\cal D}^{\ell,v}_t$ denote inverse propagators along $L_{\ell,v}$, and ${\cal J}_{A_c^\ell}$, ${\cal J}_{B_d^\ell}$, and ${\cal J}_{B^k}$ are Berends-Giele currents \cite{Berends:1987me}. The sum over “division” runs over all ways of partitioning the external legs in $\boldsymbol{A}$ and $\boldsymbol{B}$ into blocks, and the sum over $\Gamma$ ranges over diagrams consistent with these partitions and orderings.

The diagram can be divided at vertex $v$ into left and right components, allowing the sum over diagrams $\Gamma$ to be factored as
\begin{align}
	\sum_{\Gamma}=\sum_{\Gamma_{L}}\sum_{\Gamma_{R}}
\end{align}
Therefore, the amplitude becomes
\bea
{\cal A}^{{\rm Tr}(\phi^3)}_n(1,\cdots,n)=\sum_{\rm division}\,{\cal J}_{B^k}\,&&\sum_{\Gamma_L}\,\Big[\Big(\prod_{c=1}^{m_i}\,{\cal J}_{A^i_c}\Big)\,
\Big(\prod_{d=1}^{h_i}\,{\cal J}_{B^i_d}\Big)\,\Big(\prod_{t=1}^{m_i+h_i}{1\over{\cal D}^{i,v}_t}\Big)\,\Big]\nn
&&\times\sum_{\Gamma_R}\Big[\Big(\prod_{c=1}^{m_j}\,{\cal J}_{A^j_c}\Big)\,
\Big(\prod_{d=1}^{h_j}\,{\cal J}_{B^j_d}\Big)\,\Big(\prod_{t=1}^{m_j+h_j}{1\over{\cal D}^{j,v}_t}\Big)\Big]\,,~~~\label{2.15}
\eea

The key insight is that, after summing over diagrams in $\Gamma_L$ or $\Gamma_R$, the product of propagators along $L_{i,v}$ factorizes as
\bea
\sum_{\Gamma_{L}}\,\prod_{t=1}^{m_i+h_i}{1\over{\cal D}^{i,v}_t}
&=&\Big(\prod_{c=1}^{m_i}\,{1\over s_{i A^i_1\cdots A^i_c}}\Big)\,\Big(\prod_{d=1}^{h_i}\,{1\over s_{i B^i_1\cdots B^i_d}}\Big)\,,~~\label{for0-step2}
\eea
with an analogous factorization holding for the propagators along $L_{j,v}$.
This factorization in \eref{for0-step2} can be proven by induction.  The base cases for small $m_i$ and $h_i$ are readily verified. 
\begin{figure}[ht]
	\centering
	\includegraphics[width=14cm]{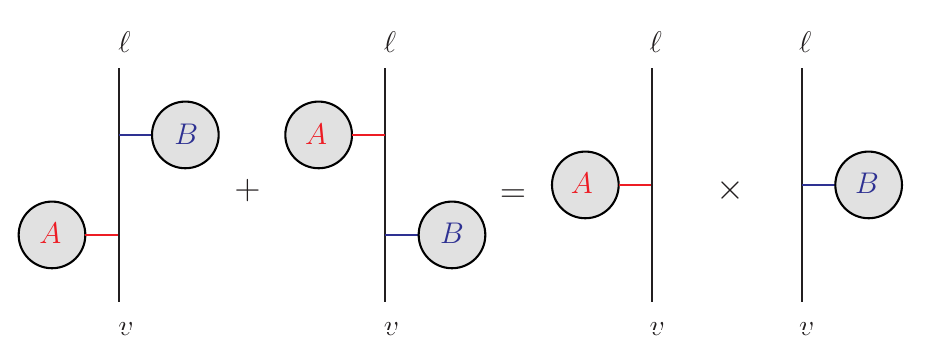} 
	\caption{Summing over diagrams with $m_{\ell}=1$, $h_{\ell}=1$.}\label{fac-exam}
\end{figure}
For instance, the factorization is trivial for $m_i+h_i=1$. The simplest non-trivial example, $m_i=h_i=1$, is shown in Fig. \ref{fac-exam}. For this case, it can be directly verified that:
\bea
\sum_{\Gamma_{L}}\,\prod_{t=1}^{2}{1\over{\cal D}^{i,v}_t}&=&{1\over s_{i B_1}}\,{1\over s_{i A_1B_1}}+{1\over s_{i A_1}}\,{1\over s_{i A_1B_1}}={1\over s_{i A_1}}\,{1\over s_{i B_1}}\,,~~\label{fac-example}
\eea
which  holds because the kinematic condition from \eqref{kinematic-condi-split-phi3} implies  $s_{i A_1} + s_{i B_1} = s_{i A_1 B_1}$.

\begin{figure}[ht]
	\centering
	\includegraphics[scale=0.8]{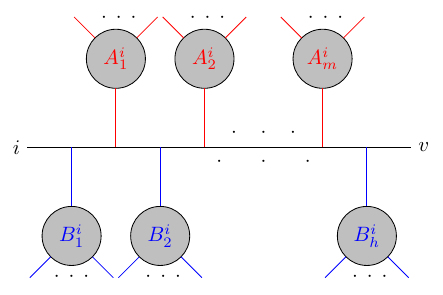}
	\includegraphics[scale=0.8]{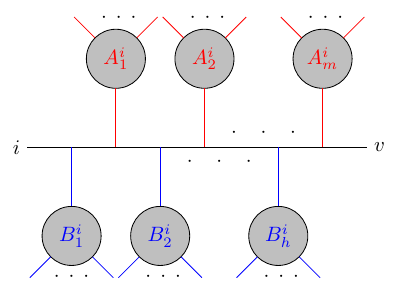}
	\caption{Shuffling blocks along $L_{i,v}$.}
	\label{facrecur}
\end{figure}
For the inductive step, assume the factorization holds for the sets $\{A_1^i,\cdots,A_m^i\}\cup\{B_1^i,\cdots,B_{h-1}^i\}$ and $\{A_1^i,\cdots,A_{m-1}^i\}\cup\{B_1^i,\cdots,B_{h}^i\}$. When adding the final block, say $B_h^i$ (or $A_m^i$), to the set $\{A_1^i,\cdots,A_m^i\}\cup\{B_1^i,\cdots,B_{h-1}^i\}$ (or $\{A_1^i,\cdots,A_{m-1}^i\}\cup\{B_1^i,\cdots,B_h^i\}$) , we consider all allowed permutations as shown in Fig. \ref{facrecur} , where $B_h^i$ may appear either to the right ($\Gamma_{L_1}$) of $A_m^i$ or to the left ($\Gamma_{L_2}$). Using the induction hypothesis, the sum over these configurations becomes:
\bea
\sum_{\Gamma_{L}}\,\prod_{t=1}^{m+h}{1\over{\cal D}^{i,v}_t}
&=&\Big(\sum_{\Gamma_{L_1}}\,\prod_{j=1}^{m+h-1}{1\over{\cal D}^{i,v}_j}\,+\,\sum_{\Gamma_{L_2}}\,\prod_{j=1}^{m+h-1}{1\over{\cal D}^{i,v}_j}\Big)\,{1\over s_{iA_1\cdots A_mB_1\cdots B_h}}\nn
&=&\Big(\prod_{b=1}^{h-1}\,{1\over s_{iB_1\cdots B_b}}\Big)\Big(\prod_{a=1}^{m-1}\,{1\over s_{iA_1\cdots A_a}}\Big)\,
\Big({1\over s_{iA_1\cdots A_m}}\Big)\,{1\over s_{iA_1\cdots A_mB_1\cdots B_h}}\nn
&&+\Big(\prod_{a=1}^{m-1}\,{1\over s_{iA_1\cdots A_a}}\Big)\Big(\prod_{b=1}^{h-1}\,{1\over s_{iB_1\cdots B_b}}\Big)\,
\Big({1\over s_{iB_1\cdots B_h}}\Big)\,{1\over s_{iA_1\cdots A_mB_1\cdots B_h}}\nn
&=&\Big(\prod_{a=1}^{m-1}\,{1\over s_{iA_1\cdots A_a}}\Big)\,\Big(\prod_{b=1}^{h-1}\,{1\over s_{iB_1\cdots B_b}}\Big)\,
\Big({1\over s_{iA_1\cdots A_m}}+{1\over s_{iB_1\cdots B_h}}\Big)\,{1\over s_{iA_1\cdots A_mB_1\cdots B_h}}\nn
&=&\Big(\prod_{a=1}^{m}\,{1\over s_{iA_1\cdots A_a}}\Big)\,\times\,\Big(\prod_{b=1}^{h}\,{1\over s_{iB_1\cdots B_b}}\Big)\,,~~\label{fac-mkb}
\eea
where the final step again uses the kinematic identity  \eqref{kinematic-condi-split-phi3}. A similar argument applies to the right component of $\Gamma$.

Substituting this factorization back into the expression \eqref{2.15} for the amplitude yields an expression that matches the $2$-split form in Eq. \eqref{zero-phi3}. The resulting amplitude is identified as:
\bea
\sum_{\text{division of A,B}} {\cal J}_{B^k} \prod_{\ell=i,j} \left[ \left( \prod_{c=1}^{m_\ell} \frac{{\cal J}_{A^\ell_c}}{s_{\ell A^\ell_1\cdots A^\ell_c}} \right) \left( \prod_{d=1}^{h_\ell} \frac{{\cal J}_{B^\ell_d}}{s_{\ell B^\ell_1\cdots B^\ell_d}} \right) \right],
\eea
which matches the expected $2$-split form \eqref{zero-phi3}, with
\begin{align}
	{\cal J}^{{\rm Tr}(\phi^3)}_{n_1}(i,\boldsymbol{A},j,\kappa) &= \sum_{\text{division of A}} \prod_{\ell=i,j} \left( \prod_{c=1}^{m_\ell} \frac{{\cal J}_{A^\ell_c}}{s_{\ell A^\ell_1\cdots A^\ell_c}} \right),\notag\\
	{\cal J}^{{\rm Tr}(\phi^3)}_{n+3-n_1}(j,\boldsymbol{B} \shuffle \kappa^*, i) &= \sum_{\text{division of B}} {\cal J}_{B^k} \prod_{\ell=i,j} \left( \prod_{d=1}^{h_\ell} \frac{{\cal J}_{B^\ell_d}}{s_{\ell B^\ell_1\cdots B^\ell_d}} \right).
\end{align}

In the above derivation, permutations of blocks within $\boldsymbol{A}$ and $\boldsymbol{B}$,  preserving the internal relative ordering within the $A$-type blocks and $B$-type blocks. For future reference, we define this operation as the shuffle of blocks from $A$ and $B$.

The extension of the above method to general BAS amplitudes is straightforward. For BAS amplitudes with different orderings $\boldsymbol{\sigma}_n \neq \boldsymbol{\sigma}'_n$, the $2$-split identity does not generally hold. However, when the orderings are compatible as defined by \eqref{compa-order}, the argument remains valid. Although fewer diagrams contribute due to ordering constraints, the surviving diagrams still yield a factorized structure consistent with \eqref{for0-step2}, thus establishing the $2$-split identities \eqref{split-BAS} and \eqref{split-BAS-2} in this case.

\section{BAS$\oplus$X}
\label{sec-BAS+X}

In Section \ref{sec-BAS}, we analyzed the $2$-split behavior of pure ${\rm Tr}(\phi^3)$/BAS amplitudes by studying only the Feynman diagrams and Feynman rules. A natural question then arises: {\bf can this diagrammatic approach be extended to a broader class of amplitudes}? In this section, we address this question by applying the method to demonstrate the $2$-split behavior of a particularly special class of  mixed BAS$\oplus$X amplitudes, where X denotes an additional theory, such as YM, GR, or NLSM. 

We first observe that the factorization \eref{for0-step2} can be applied as long as the contribution from any vertex on the line $L_{\ell,v}$---which connects a certain block (or blocks)---is unaffected when shuffling blocks from $A$ and $B$. Additionally, blocks from $A$ and $B$ are not be attached to a common vertex. This condition implies that shuffling the blocks only changes propagators along $L_{\ell,v}$, so that the effective  from summing over diagrams $\Gamma^{\ell,v}_{m_\ell,h_\ell}$ reduces to the left-hand side of \eref{for0-step2}. For example, in Fig.~\ref{fac-exam}, each diagram on the left-hand side contains two vertices: one connects to block $A$ and the other to block $B$. These two diagrams are related by a shuffle of blocks $A$ and $B$, and the contribution from each vertex remains constant $1$, unaffected by the shuffle. The feature ensures that the effective contribution from summing over these diagrams is given by the summation in \eref{fac-example}. 

However, The BAS$\oplus$X amplitudes feature nontrivial interaction vertices, unlike the pure BAS amplitudes. Therefore, we do not attempt to prove the $2$-split for general BAS$\oplus$X amplitudes using the Feynman diagram method. Instead, we consider a special case of the BAS$\oplus$X amplitudes, denoted as ${\cal A}^{\rm BAS\oplus X}_n(i_\phi,\pmb{A}_\phi,j_\phi,\pmb{B}_x\shuffle k_\phi|i_\phi,\pmb{A}'_\phi\shuffle' k_\phi,j_\phi;B_x)$ (or ${\cal A}^{{\rm BAS}\oplus{\rm X}}_n(i_\phi,\pmb{A}_x,j_\phi,\pmb{B}_\phi\shuffle k_\phi|i_\phi,\pmb{B}'_\phi\shuffle' k_\phi,j_\phi;A_x)$ for ${\rm X}={\rm YM},{\rm NLSM}$,
and ${\cal A}^{\rm BAS\oplus {\rm GR}}_n(i_\phi,\pmb{A}_\phi,j_\phi, k_\phi;B_h|i_\phi,\pmb{A}'_\phi\shuffle' k_\phi,j_\phi;B_h)$ (or ${\cal A}^{{\rm BAS}\oplus{\rm GR}}_n(i_\phi,j_\phi,\pmb{B}_\phi\shuffle k_\phi;A_h|i_\phi,\pmb{B}'_\phi\shuffle' k_\phi,j_\phi;A_h)$ for ${\rm X}={\rm GR}$, where the particles in $A\cup\{i,j,k\}$ (or $B\cup\{i,j,k\}$) are BAS scalars.
In the above notation, particles denoted by $B_x$ (or $A_x$) after semicolon correspond to $X$-type particles which do not participate in the ordering. For these particular configurations, the diagrammatic method introduced in Section \ref{sec-BAS} can be directly applied to the case ${\rm X}={\rm YM}$ case. However, for ${\rm X}={\rm GR}$ and ${\rm X}={\rm NLSM}$, a suitable extension of previous method is required.

\subsection{BAS$\oplus$YM}
\label{subsec-BAS+YM}

 We now consider the single-trace tree-level YMS amplitudes. By the minimal coupling principle,  the Lagrangian \eqref{Lag-BAS} is promoted to
 \bea
 {\cal L}_{{\rm BAS}\oplus{\rm YM}}\,=\,-{\rm tr}\Big({1\over4}\,F_{\mu\nu}F^{\mu\nu}+{1\over2}\,D_\mu\phi^A\,D^\mu\phi^A
 \Big)-{\lambda\over 3!}\,F^{ABC}f^{abc}\,\phi^{Aa}\phi^{Bb}\phi^{Cc}\,,~~\label{Lag-YS}
 \eea
 where $F^{ABC}$ and $f^{abc}$ are structure constants of flavor and gauge groups, respectively. We focus on the color/flavor-ordered partial amplitudes, with all coupling constants factored out.
 
Based on the Lagrangian \eref{Lag-YS}, we analyse the general diagrammatic structure of the target partial amplitude ${\cal A}^{{\rm BAS}\oplus{\rm YM}}_n(i_\phi,\pmb{A}_\phi,j_\phi,\pmb{B}_g\shuffle k_\phi|i_\phi,\pmb{A}'_\phi\shuffle' k_\phi,j_\phi;B_g)$. Since $i,j,k$ are scalars, all propagators along line $L_{\ell,v}$  must be scalars (see Fig. \ref{npphi3}). Consequently, the vertices connecting blocks from $B$ to $L_{\ell,v}$ are $V^{\mu}(1_{\phi},2_{\phi},3_g)\propto (k_2-k_1)^{\mu}$ and $V^{\mu\nu}(1_{\phi},2_{\phi},3_g,4_g)\propto \eta^{\mu\nu}$. The 4-vertex is clearly independent of the position of blocks from $B$. For 3-vertex, due to the constraints given by \eref{kinematic-condi-split-phi3} and  \eref{condi-polar-1}, momenta from $A$ cannot enter the block $B$. Meanwhile, the Lagrangian \eref{Lag-YS} also indicates pure scalar vertices $V(1_\phi,2_\phi,3_\phi)\propto\lambda$, 
which connect blocks from $A$ to $L_{\ell,v}$ and remain invariant under the shuffle operation. Therefore, All blocks contributions  remains unchanged under shuffling of blocks from $A$ and $B$.


Consequently, the $2$-split behavior can be extended to the BAS$\oplus$YM case, yielding the following factorization:
\bea
& &{\cal A}^{{\rm BAS}\oplus{\rm YM}}_n(i_\phi,\pmb{A}_\phi,j_\phi,\pmb{B}_g\shuffle k_\phi|i_\phi,\pmb{A}'_\phi\shuffle' k_\phi,j_\phi;B_g)\xrightarrow[]{\eref{kinematic-condi-split-phi3}\,,\eref{condi-polar-1}}\nn
& &~~~~~~~~~~~~~~~~-{\cal J}^{\rm BAS}_{n_1}(i,\pmb{A},j,\kappa|i,\pmb{A}'\shuffle'\kappa,j)\,\times\,{\cal J}_{n+3-n_1}^{{\rm Tr}(\phi^3)\oplus{\rm YM}}(j_\phi,\pmb{B}_g\shuffle\kappa^*_\phi,i_\phi)\,,~~\label{keysplit-YM-1}
\eea
or
\bea
& &{\cal A}^{{\rm BAS}\oplus{\rm YM}}_n(i_\phi,\pmb{A}_g,j_\phi,\pmb{B}_\phi\shuffle k_\phi|i_\phi,\pmb{B}'_\phi\shuffle' k_\phi,j_\phi;A_g)\xrightarrow[]{\eref{kinematic-condi-split-phi3}\,,\eref{condi-polar-1}}\nn
& &~~~~~~~~~~~~~~~~-{\cal J}^{{\rm Tr}(\phi^3)\oplus{\rm YM}}_{n_1}(i_\phi,\pmb{A}_g,j_\phi,\kappa_\phi)\,\times\,{\cal J}_{n+3-n_1}^{\rm BAS}(j,\pmb{B}\shuffle\kappa^*,i|i,\pmb{B}'\shuffle'\kappa^*,j)\,.~~\label{keysplit-YM-2}
\eea
subject to the additional polarization constraints,
\bea
\epsilon_b\cdot k_a=\epsilon_a\cdot k_b=0\,.~~\label{condi-polar-1}
\eea
The condition \eref{condi-polar-1}, together with \eref{kinematic-condi-split-phi3}, ensure that gluons in blocks $B^\ell_d$ (or $A^\ell_d$), which are connected to the line $L_{\ell,v}$, are kinematically decoupled from the momenta of the scalar particles in $A$ (or $B$). As a result, contributions from these vertices are independent of the shuffling of blocks.

\subsection{BAS$\oplus$GR and BAS$\oplus$NLSM}
\label{subsec-BAS+GR}

For BAS$\oplus$GR and BAS$\oplus$NLSM amplitudes, two new classes of interaction vertices emerge: one of the form $\phi$–$\phi$–$x$–$\cdots$–$x$, and the other of the form $\phi$–$\phi$–$\phi$–$x$–$\cdots$–$x$, where $\phi$ and $x$ denote BAS scalars and particles from theory X, respectively. These vertices potentially obstruct the key factorization identity, such as  \eqref{for0-step2} established in the previous section. Specifically, in the first class of vertices, complications arise because terms of the form $k_{\phi_1}\cdot k_{\phi_2}$ appear in the numerator, where $k_{\phi_1}$ and $k_{\phi_2}$ are the momenta of two real or virtual scalars connected to the vertex. In the second class, blocks from $A$ and $B$ would be connected in a same vertex.

\begin{figure}[ht]
	\centering
	\includegraphics[width=0.6\textwidth]{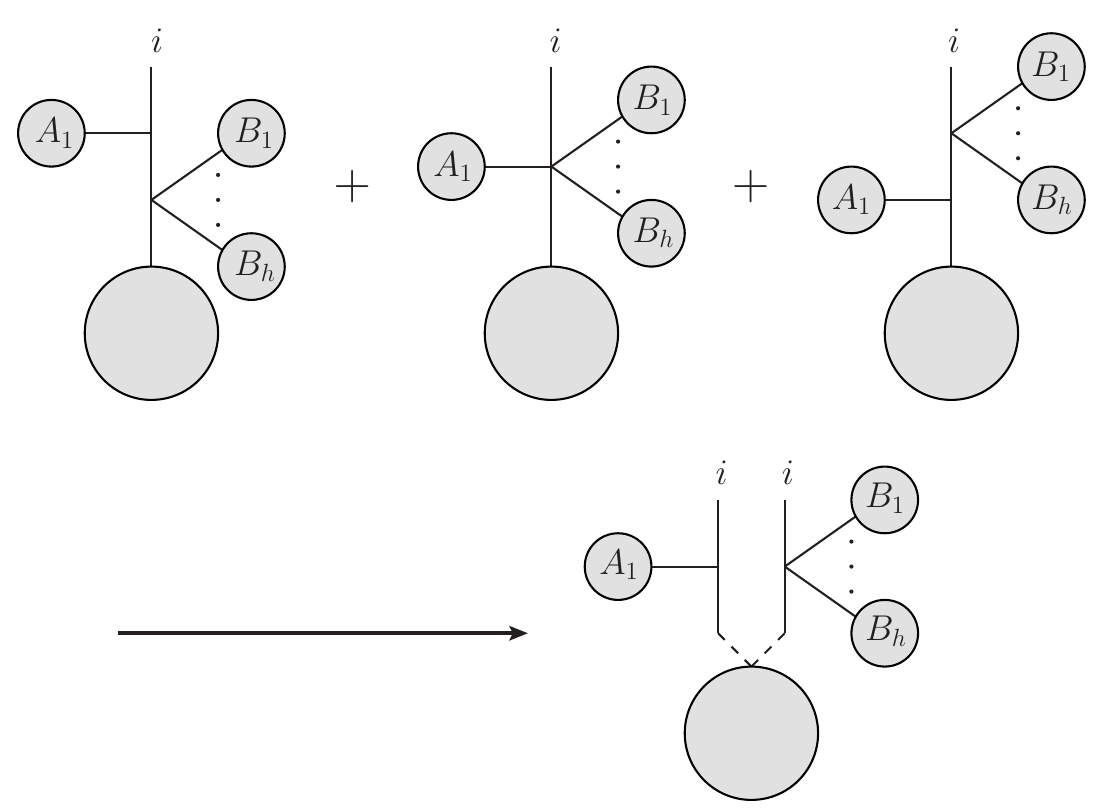}
	\caption{New shuffling and the corresponding factorization.}
	\label{shuffle3}
\end{figure}

The present subsection aims to generalize the factorization behavior \eqref{for0-step2} to incorporate these new interaction structures. The identity in \eqref{for0-step2} is based on summing over all shuffles of the blocks $A_r$ and $B_r$ on either side of a designated line $L_{\ell,v}$, while preserving the internal orderings within $\{A_r\}$ and $\{B_r\}$. Guided by a diagrammatic interpretation of "shuffling," this summation can be naturally extended, as illustrated in the first line of Fig.~\ref{shuffle3}. To isolate the contributions that potentially violate the original factorization behavior, we examine the effect of a vertex of the form $\phi$–$\phi$–$x$–$\cdots$–$x$, and decompose it as follows:
\bea
V_{\phi^2-\{B_r\}} = V'_{\phi^2-\{B_r\}} + V''_{\phi^2-\{B_r\}}=V'_{\phi^2-\{B_r\}} +\alpha_{\{B_r\}}k_{\phi_1} \cdot k_{\phi_2}\,, \label{separate-V}
\eea
where $\alpha_{\{B_r\}}$ is a Lorentz-invariant function that may depend on the momenta and polarizations of the particles in the set $\{\phi_1,\phi_2\} \cup \{B_r\}$. We assume that neither $V'_{\phi^2-\{B_r\}}$ nor $\alpha_{\{B_r\}}$ contains a $k_{\phi_1} \cdot k_{\phi_2}$ factor. Furthermore, we assume that all $\phi$–$\phi$–$\phi$–$x$–$\cdots$–$x$ vertices satisfy the condition:
\bea
V_{\phi^3-\{B_r\}} = \alpha_{\{B_r\}}\,.~~\label{assumption-V}
\eea
The condition \eqref{assumption-V} is very nontrivial and 
we will show that the cases ${\rm X}={\rm GR}$ and ${\rm X}={\rm NLSM}$ precisely satisfy the above special assumptions. Before proceeding, we first prove the following key observation. Under the assumptions in \eref{separate-V} and \eref{assumption-V}, the summation over diagrams—depicted in the first line of Fig.~\ref{shuffle3}—continues to exhibit the desired factorization behavior, analogous to that of \eqref{for0-step2}.

To begin with, let us consider the summation of the three diagrams in the first line of Fig.~\ref{shuffle3}, where $i$ denotes an on-shell, massless external BAS scalar. The corresponding sum is given by
\bea
\sum_\Gamma\,{R_\Gamma\over\prod D^{i,v}}&=&\Big(-{(k_i+k_{A_1})\cdot(k_i+k_{A_1}+k_{B_1\cdots B_h})\over s_{iA_1}}+1-{k_i\cdot(k_i+k_{B_1\cdots B_h})\over s_{iB_1\cdots B_h}}\Big)\,{\a_{\{B_r\}}\,{\cal J}_{A_1}\,\prod_{q=1}^h\,{\cal J}_{B_q}\over s_{iA_1B_1\cdots B_h}}\nn
& &+\Big({1\over s_{iA_1}}+{1\over s_{iB_1\cdots B_h}}\Big)\,{V'_{\phi^2-\{B_r\}}\,{\cal J}_{A_1}\,\prod_{q=1}^h\,{\cal J}_{B_q}\over s_{iA_1B_1\cdots B_h}}\,,
\eea
where $D^{i,v}$ denotes propagators along the line $L_{i,v}$. In this expression, the first line collects contributions from $V''_{\phi^2-\{B_r\}}=\a_{\{B_r\}}k_{\phi_1}\cdot k_{\phi_2}$ and $V_{\phi^3-\{B_r\}}=\a_{\{B_r\}}$, while the second line accounts for contributions from $V'_{\phi^2-\{B_r\}}$. The minus signs in the first line arise due to reversing the direction of the internal momenta. It is important to note that $\a_{\{B_r\}}$ and $V'_{\phi^2-\{B_r\}}$ remain invariant under permutations of $A_1$ and $\{B_r\}$, since the split kinematics ensure that momenta and polarization vectors from $B$ do not interact with $k_{A_1}$. Making use of the on-shell condition $k_i^2=0$ and  $k_{A_1}\cdot k_{B_1\cdots B_h}=0$, the expression simplifies to
\bea
\sum_\Gamma\,{R_\Gamma\over\prod D^{i,v}}&=&\Big(-{k_i\cdot(k_i+k_{B_1\cdots B_h})\over s_{iA_1}}-{k_i\cdot(k_i+k_{B_1\cdots B_h})\over s_{iB_1\cdots B_h}}\Big)\,{\a_{\{B_r\}}\,{\cal J}_{A_1}\,\prod_{q=1}^h\,{\cal J}_{B_q}\over s_{iA_1B_1\cdots B_h}}\nn
& &+\Big({1\over s_{iA_1}}+{1\over s_{iB_1\cdots B_h}}\Big)\,{V'_{\phi^2-\{B_r\}}\,{\cal J}_{A_1}\,\prod_{q=1}^h\,{\cal J}_{B_q}\over s_{iA_1B_1\cdots B_h}}\nn
&=&\Big({{\cal J}_{A_1}\over s_{iA_1}}\Big)\,\Big(\prod_{q=1}^h\,{\cal J}_{B_q}\,{V_{\phi^2-\{B_r\}}\,\over s_{iB_1\cdots B_h}}\Big)\,,~~\label{fac-shuffle3-example}
\eea
where in the final step we have used the identity
\bea
s_{iA_1}+s_{iB_1\cdots B_h}=s_{iA_1B_1\cdots B_h}\,.
\eea

The factorization in \eref{fac-shuffle3-example} can be understood as follows. The first factor corresponds to the interaction between the on-shell particle $i$ and a virtual particle carrying the momentum block $A_1$ via a cubic BAS vertex, followed by propagation with factor $1/s_{iA_1}$. The second factor similarly represents the interaction between $i$ and the virtual system composed of blocks $B_1,\cdots,B_h$, via the vertex $V_{\phi^2-\{B_r\}}$, with propagation as $1/s_{iB_1\cdots B_h}$. The above interpretation is graphically depicted in the second line of Fig.~\ref{shuffle3}.

The factorization behavior in \eref{fac-shuffle3-example} provides a natural basis for recursively deriving a more general factorization structure. Suppose the blocks $B_1,\cdots,B_h$ are partitioned into $s$ sets $\{B_r\}_1,\cdots,\{B_r\}_s$, each of which is connected to a single vertex of the form $\phi-\phi-x-\cdots-x$ or $\phi-\phi-\phi-x-\cdots-x$.
For such configurations, the sum over contributing diagrams takes the form
\bea
\sum_{\Gamma_{m,s}^{i,v}}\,{R_\Gamma\over\prod\,D^{i,v}}&=&\Big(\prod_{p=1}^m\,{\cal J}_{A_p}\,\prod_{q=1}^h\,{\cal J}_{B_q}\Big)\,\Big(\sum_{\Gamma_{m,s}^{i,v}}\,{\prod_{d=1}^s\,V_{\phi^{e_d}-\{B_r\}_d}\over \prod_{t=1}^{m+3s-\sum_{d=1}^se_d}\,D^{i,v}_t}\Big)\,,~~\label{sumshuffle3-general}
\eea
where $e_d=2\,{\rm or}\,3$, corresponding to $\phi-\phi-\{B_r\}_d$ or $\phi-\phi-\phi-\{B_r\}_d$ vertices, respectively. We will show that the summation \eref{sumshuffle3-general} factorizes as
\bea
\sum_{\Gamma_{m,s}^{i,v}}\,{R_\Gamma\over\prod\,D^{i,v}}&=&\Big(\prod_{p=1}^m\,{\cal J}_{A_p}\,\prod_{c=1}^m\,{1\over s_{iA_1\cdots A_c}}\Big)\,\Big(\prod_{q=1}^h\,{\cal J}_{B_q}\,\prod_{d=1}^s\,{V_{\phi^2-\{B_r\}_d}\over s_{i\{B_r\}_1\cdots\{B_r\}_d}}\Big)\,,~~\label{fac-shuffle3-general}
\eea
where the interpretation of each resulting piece is similar as that for \eref{fac-shuffle3-example}. To prove the general factorization in \eref{fac-shuffle3-general}, we proceed by induction. We assume that \eref{fac-shuffle3-general} is valid for $\Gamma_{m-1,s}^{i,v}$, $\Gamma_{m,s-1}^{i,v}$ and $\Gamma_{m-1,s-1}^{i,v}$. For the inductive step, the summation from  \eqref{sumshuffle3-general} is decomposed into four parts:
\bea
\sum_{\Gamma_{m,s}^{i,v}}\,{\prod_{d=1}^s\,V_{\phi^{e_d}-\{B_r\}_d}\over \prod_{t=1}^{m+3s-\sum_{d=1}^se_d}\,D^{i,v}_t}&=&
\sum_{\Gamma_{m-1,s}^{i,v}}\,{\prod_{d=1}^s\,V_{\phi^{e_d}-\{B_r\}_d}\over \prod_{t=1}^{m-1+3s-\sum_{d=1}^se_d}\,D^{i,v}_t}\,{1\over s_{iA_1\cdots A_m\{B_r\}_1\cdots\{B_r\}_s}}\nn
& &+\sum_{\Gamma_{m-1,s-1}^{i,v}}\,{\prod_{d=1}^{s-1}\,V_{\phi^{e_d}-\{B_r\}_d}\over \prod_{t=1}^{m-1+3(s-1)-\sum_{d=1}^{s-1}e_d}\,D^{i,v}_t}\,{V_{\phi^3-\{B_r\}_s}\over s_{iA_1\cdots A_m\{B_r\}_1\cdots\{B_r\}_s}}\nn
& &+\sum_{\Gamma_{m,s-1}^{i,v}}\,{\prod_{d=1}^{s-1}\,V_{\phi^{e_d}-\{B_r\}_d}\over \prod_{t=1}^{m+3(s-1)-\sum_{d=1}^{s-1}e_d}\,D^{i,v}_t}\,{V''_{\phi^2-\{B_r\}_s}\over s_{iA_1\cdots A_m\{B_r\}_1\cdots\{B_r\}_s}}\nn
& &+\sum_{\Gamma_{m,s-1}^{i,v}}\,{\prod_{d=1}^{s-1}\,V_{\phi^{e_d}-\{B_r\}_d}\over \prod_{t=1}^{m+3(s-1)-\sum_{d=1}^{s-1}e_d}\,D^{i,v}_t}\,{V'_{\phi^2-\{B_r\}_s}\over s_{iA_1\cdots A_m\{B_r\}_1\cdots\{B_r\}_s}}\,.~~\label{divide-to3}
\eea
Applying the inductive hypothesis to each parts and using the assumptions \eref{separate-V} and \eref{assumption-V} for vertices $V''_{\phi^2-\{B_r\}_s}$ and $V_{\phi^3-\{B_r\}_s}$, we evaluate the four terms as follows:
\bea
& &\textbf{(1) The fisrt part:}\nn
& &\sum_{\Gamma_{m-1,s}^{i,v}}\,{\prod_{d=1}^s\,V_{\phi^{e_d}-\{B_r\}_d}\over \prod_{t=1}^{m-1+3s-\sum_{d=1}^se_d}\,D^{i,v}_t}\,{1\over s_{iA_1\cdots A_m\{B_r\}_1\cdots\{B_r\}_s}}\nn
&=&\Big(\prod_{c=1}^{m-1}\,{1\over s_{iA_1\cdots A_c}}\Big)\,\Big(\prod_{d=1}^{s-1}\,{V_{\phi^2-\{B_r\}_d}\over s_{i\{B_r\}_1\cdots\{B_r\}_d}}\Big)\,{-\a_{\{B_r\}_s}\,k_{i\{B_r\}_1\cdots\{B_r\}_{s-1}}\cdot k_{i\{B_r\}_1\cdots\{B_r\}_{s}}+V'_{\phi^2-\{B_r\}_s}\over s_{i\{B_r\}_1\cdots\{B_r\}_s}~~s_{iA_1\cdots A_m\{B_r\}_1\cdots\{B_r\}_s}}\,,\nn
& &\textbf{(2) The second part:}\nn
& &\sum_{\Gamma_{m-1,s-1}^{i,v}}\,{\prod_{d=1}^{s-1}\,V_{\phi^{e_d}-\{B_r\}_d}\over \prod_{t=1}^{m-1+3(s-1)-\sum_{d=1}^{s-1}e_d}\,D^{i,v}_t}\,{V_{\phi^3-\{B_r\}_s}\over s_{iA_1\cdots A_m\{B_r\}_1\cdots\{B_r\}_s}}\nn
&=&\Big(\prod_{c=1}^{m-1}\,{1\over s_{iA_1\cdots A_c}}\Big)\,\Big(\prod_{d=1}^{s-1}\,{V_{\phi^2-\{B_r\}_d}\over s_{i\{B_r\}_1\cdots\{B_r\}_d}}\Big)\,{\a_{\{B_r\}_s}\over s_{iA_1\cdots A_m\{B_r\}_1\cdots\{B_r\}_s}}\,,\nn
& &\textbf{(3) The third part:}\nn
& &\sum_{\Gamma_{m,s-1}^{i,v}}\,{\prod_{d=1}^{s-1}\,V_{\phi^{e_d}-\{B_r\}_d}\over \prod_{t=1}^{m+3(s-1)-\sum_{d=1}^{s-1}e_d}\,D^{i,v}_t}\,{V''_{\phi^2-\{B_r\}_s}\over s_{iA_1\cdots A_m\{B_r\}_1\cdots\{B_r\}_s}}\nn
&=&\Big(\prod_{c=1}^{m-1}\,{1\over s_{iA_1\cdots A_c}}\Big)\,\Big(\prod_{d=1}^{s-1}\,{V_{\phi^2-\{B_r\}_d}\over s_{i\{B_r\}_1\cdots\{B_r\}_d}}\Big)\,{-\a_{\{B_r\}_s}\,k_{iA_1\cdots A_h\{B_r\}_1\cdots\{B_r\}_{s-1}}\cdot k_{iA_1\cdots A_m\{B_r\}_1\cdots\{B_r\}_{s}}\over s_{iA_1\cdots A_m}~~s_{iA_1\cdots A_m\{B_r\}_1\cdots\{B_r\}_s}}\,,\nn
& &\textbf{(4) The forth part:}\nn
& &\sum_{\Gamma_{m,s-1}^{i,v}}\,{\prod_{d=1}^{s-1}\,V_{\phi^{e_d}-\{B_r\}_d}\over \prod_{t=1}^{m+3(s-1)-\sum_{d=1}^{s-1}e_d}\,D^{i,v}_t}\,{V'_{\phi^2-\{B_r\}_s}\over s_{iA_1\cdots A_m\{B_r\}_1\cdots\{B_r\}_s}}\nn
&=&\Big(\prod_{c=1}^{m-1}\,{1\over s_{iA_1\cdots A_c}}\Big)\,\Big(\prod_{d=1}^{s-1}\,{V_{\phi^2-\{B_r\}_d}\over s_{i\{B_r\}_1\cdots\{B_r\}_d}}\Big)\,{V'_{\phi^2-\{B_r\}_s}\over s_{iA_1\cdots A_m}~~s_{iA_1\cdots A_m\{B_r\}_1\cdots\{B_r\}_s}}\,.~~\label{evaluate-3part}
\eea
Plugging \eref{evaluate-3part} back into \eref{divide-to3}, and using $k_{A_1\cdots A_m}\cdot k_{\{B_r\}_1\cdots\{B_r\}_{s-1}}=0$, $k_{A_1\cdots A_m}\cdot k_{\{B_r\}_1\cdots\{B_r\}_{s}}=0$,
as well as $s_{iA_1\cdots A_m}=k_{iA_1\cdots A_m}^2$, we find
\bea
& &\sum_{\Gamma_{m,s}^{i,v}}\,{\prod_{d=1}^s\,V_{\phi^{e_d}-\{B_r\}_d}\over \prod_{t=1}^{m+3s-\sum_{d=1}^se_d}\,D^{i,v}_t}\nn
&=&
\Big(\prod_{c=1}^{m-1}\,{1\over s_{iA_1\cdots A_c}}\Big)\,\Big(\prod_{d=1}^{s-1}\,{V_{\phi^2-\{B_r\}_d}\over s_{i\{B_r\}_1\cdots\{B_r\}_d}}\Big)\,\big(-\a_{\{B_r\}_s}\,k_{i\{B_r\}_1\cdots\{B_r\}_{s-1}}\cdot k_{i\{B_r\}_1\cdots\{B_r\}_{s}}+V'_{\phi^2-\{B_r\}_s}\big)\nn
& &\Big({1\over s_{iA_1\cdots A_m}}+{1\over s_{i\{B_r\}_1\cdots\{B_r\}_s}}\Big)\,{1\over s_{iA_1\cdots A_m\{B_r\}_1\cdots\{B_r\}_s}}\nn
&=&\Big(\prod_{c=1}^{m}\,{1\over s_{iA_1\cdots A_c}}\Big)\,\Big(\prod_{d=1}^{s}\,{V_{\phi^2-\{B_r\}_d}\over s_{i\{B_r\}_1\cdots\{B_r\}_d}}\Big)\,,
\eea
where the last step used
\bea
s_{iA_1\cdots A_m}+s_{i\{B_r\}_1\cdots\{B_r\}_s}=s_{iA_1\cdots A_m\{B_r\}_1\cdots\{B_r\}_s}\,.
\eea
The above result turns the summation \eref{sumshuffle3-general} into the desired factorization formula \eref{fac-shuffle3-general}.

Now we consider the application of the above discussion to GR. Firstly, since GR amplitudes are unordered, the block $B$ can be placed anywhere between $i$ and $j$, as illustrated in Fig. \ref{shuffle3}. Once the block $B$ and the relative ordering are fixed, however, the previous analysis can be applied straightforwardly.
Secondly, the factorization property \eref{fac-shuffle3-general} is based on the assumption that the $V''_{\phi^2-\{B_r\}}$ and $V_{\phi^3-\{B_r\}}$ vertices share a common $\a_{\{B_r\}}$. For GR, this condition is indeed satisfied. This can be seen from the action:
\bea
S=\int\,{d^d x}\,\sqrt{-g}\,\Big[{1\over2}\, g_{\mu\nu}\partial^\mu\phi^{Aa}\,\partial^{\nu}\phi^{Aa}+{\lambda\over3!}\,F^{ABC}f^{abc}\,
\phi^{Aa}\phi^{Bb}\phi^{Cc}\Big]\,.
\eea
The vertex $V_{\phi^2-\{B_r\}_h}$ (with $h$ denoting a graviton) originates from the term $\sqrt{-g}g_{\mu\nu}\partial^\mu\phi\partial^\nu\phi$. Expanding the metric as $g_{\mu\nu}=\eta_{\mu\nu}+\kappa h_{\mu\nu}$, the part $\eta_{\mu\nu}$ yields $k_{\phi_1}\cdot k_{\phi_2}$. Consequently, both the vertices $V''_{\phi^2-\{B_r\}_h}$ and $V_{\phi^3-\{B_r\}_h}$ arise from the expansion of the overall prefactor $\sqrt{-g}$, which precisely corresponds to condition \eqref{assumption-V}. The part $h_{\mu\nu}\partial^\mu\phi\partial^\nu\phi$  contributes to the $V'$ part in \eref{separate-V}. 

Since GR satisfies above pattern, we utilize the factorization \eref{fac-shuffle3-general} to write down the $2$-split of mixed amplitudes in BAS$\oplus$GR. Let us first focus on the BAS$\oplus$GR amplitude ${\cal A}^{{\rm BAS}\oplus{\rm GR}}(i_\phi,\pmb A_\phi\shuffle k_\phi,j_\phi;B_h|i_\phi,\pmb A'_\phi\shuffle' k_\phi,j_\phi;B_h)$, which can be expressed as
\bea
& &{\cal A}^{{\rm BAS}\oplus{\rm GR}}(i_\phi,\pmb A_\phi\shuffle k_\phi,j_\phi;B_h|i_\phi,\pmb A'_\phi\shuffle' k_\phi,j_\phi;B_h)\nn
&=&\sum_{\rm division}\,{\cal J}_{B^k}\,\prod_{\ell=i,j}\,\Big[\,\Big(\prod_{p=1}^{m_\ell}\,{\cal J}_{A^\ell_p}\Big)\,
\Big(\prod_{q=1}^{h_\ell}\,{\cal J}_{B^\ell_q}\Big)\,\Big(\sum_{\Gamma_{m,s}^{\ell,v}}\,{\prod_{d=1}^{s_\ell}\,V_{\phi^{e_d}-\{B_r\}_d}\over \prod_{t=1}^{m+3s_\ell-\sum_{d=1}^{s_\ell}e_d}\,D^{\ell,v}_t}\Big)\,\Big]\,.~~\label{np-GR+BAS}
\eea
Substituting the factorization behavior in \eref{fac-shuffle3-general}, we obtain
\bea\label{keysplit-BAS+GR} 
& &{\cal A}^{{\rm BAS}\oplus{\rm GR}}(i_\phi,\pmb A_\phi\shuffle k_\phi,j_\phi;B_h|i_\phi,\pmb A'_\phi\shuffle' k_\phi,j_\phi;B_h)\,\xrightarrow[]{\eref{kinematic-condi-split-phi3}\,,\,{\eref{condi-polar-2}}}\nn
& &~~~~~~~~~~~~~~~~~~~~~~~~~~~~~~~~{\cal J}^{\rm BAS}_{n_1}(i,\pmb A\shuffle \kappa,j|i,\pmb A'\shuffle' \kappa,j)\,\times\,
{\cal J}_{n+3-n_1}^{{\rm Tr}(\phi^3)\oplus{\rm GR}}(j_\phi,i_\phi,\kappa^*_\phi;B_h)\,,
\eea
where
\bea
{\cal J}^{\rm BAS}_{n_1}(i,\pmb A\shuffle \kappa,j|i,\pmb A'\shuffle' \kappa,j)&=&\sum_{\rm division}\,\prod_{\ell=i,j}\,\Big[\,\Big(\prod_{p=1}^{m_\ell}\,{\cal J}_{A^\ell_p}\Big)\,
\Big(\prod_{p=1}^{m_\ell}\,{1\over s_{\ell A^\ell_1\cdots A^\ell_p}}\Big)\,\Big]\,,\nn
{\cal J}_{n+3-n_1}^{{\rm Tr}(\phi^3)\oplus{\rm GR}}(j_\phi,i_\phi,\kappa^*_\phi;B_h)&=&\sum_{\rm division}\,{\cal J}_{B^k}\,\prod_{\ell=i,j}\,\Big[\Big(\prod_{q=1}^{h_\ell}\,{\cal J}_{B^\ell_q}\Big)\,
\Big(\prod_{d=1}^{s_\ell}\,{V_{\phi^2-\{B_r\}^\ell_d}\over s_{\ell \{B_r\}^\ell_1\cdots \{B_r\}^\ell_d}}\Big)\,\Big]\,.
\eea
Here, the constrains on polarization tensors of gravitons are
\bea
\epsilon_b\cdot k_a=\W\epsilon_b\cdot k_a=0\,,~~\label{condi-polar-2}
\eea
where each polarization tensor is decomposed as $\epsilon_b^{\mu\nu}=\epsilon_b^\mu\W\epsilon_b^\nu$.

For the ${\rm X}={\rm NLSM}$ case, the situation is more subtle—and correspondingly more interesting—due to the large freedom in reparameterizing the Lagrangian. From the perspective of Feynman rules, the origin of such reparameterizations is clear: each vertex $V_{\{p\}}$ or $V_{\phi^2-\{p\}}$, where the subscript $p$ denotes pions, carries mass dimension $2$. This indicates the potential presence of terms that may cancel the propagator attached to the vertex. Such cancellations allow for numerous ways to reorganize the Feynman rules. For our purposes, it suffices to identify a parameterization that satisfies the assumptions \eqref{separate-V} and \eqref{assumption-V}.
	
A suitable choice is the Lagrangian given in equation (3.36) of \cite{Mizera:2018jbh}, which is obtained from the BAS$\oplus$YM Lagrangian \eqref{Lag-YS} via a carefully designed dimensional reduction. In this formulation, neither a $k_{\phi_1}\cdot k_{\phi_2}$ term appears in $V_{\phi^2-\{p\}}$, nor does the vertex $V_{\phi^3-\{p\}}$ exist, i.e., $\a_{B_r}=0$. Consequently, the original factorization behavior expressed in \eqref{for0-step2} remains applicable. Using this parameterization, the derivation of the $2$-split proceeds in close analogy with the BAS$\oplus$YM case, as expected given the connection between the two theories via dimensional reduction.

This Lagrangian also includes a four-point vertex $4$-point $V(1_\phi,2_\phi,3_\phi,4_\phi)$ characterized solely by a coupling constant. The vertex excluded by imposing the mass dimension constrains on BAS$\oplus$NLSM amplitudes in \eref{keysplit-1} and \eref{keysplit-2}, requiring them to be $2|B_p|-2(n-3)$ and $2|A_p|-2(n-3)$, respectively. This constraint again enables the interpretation of one of the two resulting currents in the $2$-split as the pure BAS current.

An alternative choice is to adopt the method of \cite{Cachazo:2016njl}, which defines both $V_{\phi^2-\{p\}}$ and $V_{\phi^3-\{p\}}$ vertices. The explicit forms of these vertices was shown in \cite{Low:2017mlh, Yin:2018hht}, we now use them to verify our assumptions \eref{separate-V} and \eref{assumption-V}. For BAS$\oplus$NLSM apmlitudes involving $n-2$ NLSM particles and $2$ $\phi$'s, the Feynman rules coincide with those of an $n$-point NLSM amplitude. The $2n$-point vertex in the exponential parameterization \cite{Yin:2018hht, Kampf:2013vha} is
\begin{align}\label{eq:2phiver}
	V_{2n}(p_1,\ldots,p_n)=\frac{(-1)^n}{(2n)!}\left(\frac{4}{F^2}\right)^{n-1}\sum_{k=1}^{2n-1}(-1)^{k-1}{{2n-2}\choose{k-1}}\sum_{i=1}^{2n}(k_i\cdot k_{i+k})\,,
\end{align}
where $k_{2n+i}\equiv k_i$. The $(2n+1)$-point vertex, which involves 3 $\phi$'s---two of which are adjacent---is \cite{Low:2017mlh, Yin:2018hht}
\begin{align}\label{eq:3phiver}
	V^{\rm NLSM\oplus\phi^3}(\mathbb{I}_{2n+1}|1,2n+1,j)=\frac{1}{2}\frac{-(-4)^n\lambda}{(2n+1)!F^{2n-2}}\left[{{2n}\choose{j-1}}(-1)^{j-1}-1\right]\,.
\end{align}
In order for \eqref{fac-shuffle3-general} to be valid in BAS$\oplus$NLSM, the $V''_{\phi^2-\{B_r\}}$ and $V_{\phi^3-\{B_r\}}$ vertices must share a common $\a_{\{B_r\}}$. Since the particles on the line $L_{\ell,v}$ are $\phi$'s, the 2 $\phi$'s in the vertex $V_{\phi^2-\{B_r\}_p}$ are adjacent. Consequently, for the term $k_{\phi_1}\cdot k_{\phi_2}$ to appear in \eqref{eq:2phiver}, the summation index $k$  must be either $1$ or $2n-1$, yielding
\begin{align}
	\alpha_{B_r}=\frac{(-1)^n}{(2n)!}4^{n-1}(1+1)=\frac{1}{2}\frac{(-4)^n}{(2n)!}\,.
\end{align}
For vertices $V_{\phi^3-\{B_r\}_p}$, three $\phi$'s are adjacent, implying $j=2n$ in \eqref{eq:3phiver}. Thus,
\begin{align}
	V_{\phi^3-\{B_r\}_p}=\frac{1}{2}\frac{-(-4)^n}{(2n+1)!}\left[-2n-1\right]=\frac{1}{2}\frac{(-4)^n}{(2n)!}\,.
\end{align}
Clearly, $V_{\phi^3-\{B_r\}_p}=\alpha_{B_r}$. Thus, the derivation of $2$-split for BAS$\oplus$GR amplitudes can be drectly applied to the current BAS$\oplus$NLSM case.

Consequently, either of two parameterizations lead to the following $2$-split behavior of the BAS$\oplus$NLSM amplitude
\bea
& &{\cal A}^{{\rm BAS}\oplus{\rm NLSM}}_n(i_\phi,\pmb{A}_\phi,j_\phi,\pmb{B}_p\shuffle k_\phi|i_\phi,\pmb{A}'_\phi\shuffle' k_\phi,j_\phi;B_p)\,\xrightarrow[]{\eref{kinematic-condi-split-phi3}}\nn
& &~~~~~~~~~~~~~~~~~~~~{\cal J}^{\rm BAS}_{n_1}(i,\pmb{A},j,\kappa|i,\pmb{A}'\shuffle'\kappa,j)\,\times\,{\cal J}_{n+3-n_1}^{{\rm BAS}\oplus{\rm NLSM}}(j_\phi,\pmb{B}_p\shuffle\kappa^*_\phi,i_\phi|i_\phi,\kappa^*_\phi,j_\phi;B_p)\nn
& &~~~~~~~~~~~~~~~~=-{\cal J}^{\rm BAS}_{n_1}(i,\pmb{A},j,\kappa|i,\pmb{A}'\shuffle'\kappa,j)\,\times\,{\cal J}_{n+3-n_1}^{{\rm Tr}(\phi^3)\oplus{\rm NLSM}}(j_\phi,\pmb{B}_p\shuffle\kappa^*_\phi,i_\phi)\,,~~\label{keysplit-1}
\eea
where the final step has used the ordered reversed relation
\bea
{\cal J}_{n+3-n_1}^{{\rm BAS}\oplus{\rm NLSM}}(j_\phi,\pmb{B}_p\shuffle\kappa^*_\phi,i_\phi|i_\phi,\kappa^*_\phi,j_\phi;B_p)=-{\cal J}_{n+3-n_1}^{{\rm BAS}\oplus{\rm NLSM}}(j_\phi,\pmb{B}_p\shuffle\kappa^*_\phi,i_\phi|j_\phi,\kappa^*_\phi,i_\phi;B_p)\,,
\eea
along with the observation
\bea
{\cal J}_{n+3-n_1}^{{\rm BAS}\oplus{\rm NLSM}}(j_\phi,\pmb{B}_p\shuffle\kappa^*_\phi,i_\phi|j_\phi,\kappa^*_\phi,i_\phi;B_p)={\cal J}_{n+3-n_1}^{{\rm Tr}(\phi^3)\oplus{\rm NLSM}}(j_\phi,\pmb{B}_p\shuffle\kappa^*_\phi,i_\phi)\,.~~\label{obser-phi-BAS}
\eea
Similarly, if $i$, $j$, $k$ and $b\in B$ are BAS scalars, we obtain the analogous $2$-split,
\bea
& &{\cal A}^{{\rm BAS}\oplus{\rm NLSM}}_n(i_\phi,\pmb{A}_p,j_\phi,\pmb{B}_\phi\shuffle k_\phi|i_\phi,\pmb{B}'_\phi\shuffle' k_\phi,j_\phi;A_p)\,\xrightarrow[]{\eref{kinematic-condi-split-phi3}}\nn
& &~~~~~~~~~~~~~~~~-{\cal J}^{{\rm Tr}(\phi^3)\oplus{\rm NLSM}}_{n_1}(i_\phi,\pmb{A}_p,j_\phi,\kappa_\phi)\,\times\,{\cal J}_{n+3-n_1}^{\rm BAS}(j,\pmb{B}\shuffle\kappa^*,i|i,\pmb{B}'\shuffle'\kappa^*,j)\,.~~\label{keysplit-2}
\eea

\section{Review of universal expansions of tree-level amplitudes}
\label{sec-expansion}

In the previous two sections, we extended the diagrammatic method introduced in \cite{Zhou:2024ddy} and applied it to prove the $2$-split behavior for a special class of BAS$\oplus$X amplitudes. However, generalizing this approach to pure X amplitudes—particularly for theories with infinitely many distinct vertices, such as NLSM and GR—remains challenging. As discussed in Section \ref{sec-intro}, the $2$-split behavior of these pure X amplitudes can nonetheless be accessed by combining the $2$-split results for BAS$\oplus$X amplitudes derived earlier with another key property of pure X amplitudes known as universal expansions. In this section, we provide a concise review of these universal expansions.

As demonstrated in a broad range of studies \cite{Stieberger:2016lng, Schlotterer:2016cxa, Chiodaroli:2017ngp, Nandan:2016pya, delaCruz:2016gnm, Fu:2017uzt, Teng:2017tbo, Du:2017kpo, Du:2017gnh, Feng:2019tvb, Zhou:2019gtk, Zhou:2019mbe, Wei:2023yfy, Hu:2023lso, Du:2024dwm, Zhou:2024qwm, Zhou:2024qjh},
tree-level amplitudes in a wide variety of theories can be systematically expanded in terms of those in the BAS theory. The expansions of YM and gravity GR amplitudes are given by
\bea
{\cal A}^{\rm YM}_n(\pmb\sigma_n)=\sum{\pmb\a_{n-2}}C_{\rm Y}^\epsilon(\pmb\a_{n-2}){\cal A}^{\rm BAS}_n(\pmb\sigma_n|i,\pmb\a_{n-2},j)\,,
\label{expan-YM}
\eea
and
\bea
{\cal A}^{\rm GR}_n=\sum{\pmb\a_{n-2}}\sum_{\pmb\b_{n-2}}C_{\rm Y}^\epsilon(\pmb\a_{n-2})C_{\rm Y}^{\W\epsilon}(\pmb\b_{n-2})
{\cal A}^{\rm BAS}_n(i,\pmb\a_{n-2},j|i,\pmb\b_{n-2},j)\,,
\label{expan-GR}
\eea
where $\pmb\a_{n-2}$ and $\pmb\b_{n-2}$ are ordered subsets of the $n-2$ elements in $\{1,\ldots,n\}\setminus\{i,j\}$. The labels $i$ and $j$, which are fixed at the two ends of the ordering, can be chosen arbitrarily. In the context of $2$-split factorization, we will choose them to coincide with the $i$ and $j$ in \eqref{split-BAS}.

As noted at the end of the previous section, the polarization tensor of an external graviton is decomposed as $\varepsilon_q^{\mu\nu}=\epsilon_q^\mu\W\epsilon_q^\nu$, where $\epsilon_q^\mu$ and $\W\epsilon_q^\mu$ are two polarization vectors. For pure Einstein gravity, these two vectors are identical, whereas in the extended model involving Einstein gravity coupled to a $B$-field and a dilaton, they are independent.

In \eqref{expan-GR}, each kinematic coefficient $C_{\rm Y}^\epsilon(\pmb\a_{n-2})$ depends only on the polarization vectors $\epsilon_q$, while each $C_{\rm Y}^{\W\epsilon}(\pmb\b_{n-2})$ depends only on $\W\epsilon_q$, for all $q\in\{1,\ldots,n\}$. By combining \eqref{expan-GR} and \eqref{expan-YM}, the expansion of GR amplitudes in terms of YM amplitudes becomes
\bea
{\cal A}^{\rm GR}_n=\sum_{\pmb\b_{n-2}}C_{\rm Y}^{\W\epsilon}(\pmb\b_{n-2})
{\cal A}^{\rm YM}_n(i,\pmb\b{n-2},j)\,.
\label{expan-GRtoYM}
\eea

Before presenting the expansion coefficients $C_{\rm Y}^\epsilon(\pmb\a_{n-2})$, we first introduce some preparatory definitions. We define a {\bf global rerefence ordering } $\pmb{\cal G}=\{g_1,...,g_{n-2}\}$. Given another ordering
$\pmb\a_{n-2}=\{\a_1,\cdots,\a_{n-2}\}$, we define its compatible spliting (with respect to  the ordering $\pmb{\cal G}$) as the ordering of sets $\{ \pmb r_0, \pmb r_1,...,\pmb r_k\}$, satisfying the following conditions:
\begin{itemize}
	\item Each set $\pmb r_i$ is a subset of $\pmb\a_{n-2}$,  preserving the relative order of elements as in $\pmb\a_{n-2}$.
	In particular, the set $\pmb r_0$ is allowed to be empty.
	\item The subsets are disjoint and collectively exhaustive: $\pmb r_i\cap \pmb r_j=\emptyset,~~\forall i\neq j$ and $\cup_{t=0}^k \pmb r_t=\pmb\a_{n-2}$.
	\item For each set $\pmb r_i$ with $i=1,...,k$, the last element must be the leftmost element (with respect to the ordering $\pmb{\cal G}$) in the set  $\pmb{\cal G}\setminus \cup_{t=0}^{i-1}  r_t$, where $r_t$ denotes the unordered version of $\pmb r_t$, obtained by removing its internal ordering.
\end{itemize}
%

Given a compatible splitting $\{ \pmb r_0, \pmb r_1,...,\pmb r_k\}$, we define the corresponding factor as follows.
For $\pmb r_0=\{r^0_1,\cdots,r^0_{|0|}\}$ (where subscript $|k|$ stands for the length of the set $\pmb r_k$), we treat it as forming the ordered list $\{i, r^0_1,\cdots,r^0_{|0|}, j\}$. The associated kinematic factor is given by
\bea
F_{\pmb r_0}=\epsilon_j\cdot f_{r^0_{|0|}}\cdot f_{r^0_{|0|-1}}\cdots f_{r^0_1}\cdot\epsilon_i\,,~~\label{F0}
\eea
where $f_q^{\mu\nu}\equiv k_q^\mu\epsilon_q^\nu-\epsilon_q^\mu k_q^\nu$, and $\epsilon_q^\mu$ is the polarization vector for the external particle $q$. Note that  even if $\pmb r_0$ is an empty set, \eqref{F0} remains well defined.  For any ordered set $\pmb r_k$ with $k\neq 0$, the corresponding kinematic factor is defined as
\bea
F_{\pmb r_k}=\epsilon_{r^k_{|k|}}\cdot f_{r^k_{|k|-1}}\cdots f_{r^k_1}\cdot Z_{r^k_1}\,.~~\label{Fl}
\eea
Here, the combinatorial momentum is defined as 
\begin{align}
	Z_{r^k_1}=k_i+\W Z_{r^k_1}\,,
\end{align}
where $\W Z_{r^k_1}$ is the sum of momenta of external legs satisfy two conditions: (1) they appear to the left of $r^k_1$ in the ordering $\pmb\a_{n-2}$, (2) they belong to $\pmb r_h$ with $h<k$. Consequently, the coefficient $C_{\rm Y}^\epsilon(\pmb\a_{n-2})$ is constructed as
\bea
C_{\rm Y}^\epsilon(\pmb\a_{n-2})=\sum_{S[\pmb\a_{n-2}]}\,\prod_{k=0}^f\,F_{\pmb r_k}\,,~~\label{c-YM}
\eea
where $\sum_{S[\pmb\a_{n-2}]}$ denotes the sum over ordered splittings with respect to the ordering $\pmb\a_{n-2}$.

To illustrate this explicitly, consider the case $C_{\rm Y}^\epsilon(\{3,2\})$, where we take $i=1$ and $j=4$. The global reference ordering is chosen as $\mathcal{G}=\{3,2\}$. The possible candidates for $\pmb r_0$ compatible with $\mathcal{G}$ are $\emptyset$, $\{2\}$, $\{3\}$, $\{3,2\}$. For $\pmb r_0=\emptyset$, the lowest element in the global reference ordering $\mathcal{G}$ is $3$, thus $\pmb r_1=\{3\}$, which leads to the ordered splitting $\{\{1,4\},\{3\},\{2\}\}$. Similarly, for remaining three choices of $\pmb r_0$, one obtains $\{\{1,2,4\},\{3\}\}$, $\{\{1,3,4\},\{2\}\}$ and $\{\{1,3,2,4\}\}$, respectively. The coefficient $C_{\rm Y}^\epsilon(\{3,2\})$ is then determined by substituting kinematic factors from \eref{F0} and \eref{Fl},
\bea
C_{\rm Y}^\epsilon(3,2)=(\epsilon_4\cdot\epsilon_1)(\epsilon_3\cdot k_1)(\epsilon_2\cdot k_{13})+(\epsilon_4\cdot f_2\cdot\epsilon_1)(\epsilon_3\cdot k_1)+(\epsilon_4\cdot f_3\cdot\epsilon_1)(\epsilon_2\cdot k_{13})+(\epsilon_4\cdot f_2\cdot f_3\cdot\epsilon_1)\,.
\eea

The BAS$\oplus$YM and BAS$\oplus$GR amplitudes have the following analogous expansion formulas
\bea
{\cal A}^{{\rm BAS}\oplus{\rm YM}}(\pmb\sigma_n|i_\phi,\pmb\a_{\phi;m-2},j_\phi;\b_{g;n-m})&=&\sum_{\pmb\b_{n-m}}\,\sum_{\shuffle}\,C^{\epsilon}_{\rm YS}(\pmb\a_{m-2}\shuffle\pmb\b_{n-m})\,{\cal A}^{\rm BAS}(\pmb\sigma_n|i,\pmb\a_{m-2}\shuffle\pmb\b_{n-m},j)\,,~~\label{expan-YM+S}
\eea
\bea
& &{\cal A}^{{\rm BAS}\oplus{\rm GR}}(i_\phi,\pmb\a_{\phi;m-2},j_\phi;\b_{h;n-m}|i_\phi,\pmb\a'_{\phi;m-2},j_\phi;\b_{h;n-m})\nn
&=&\sum_{\pmb\b_{n-m}}\,\sum_{\pmb\b'_{n-m}}\,\sum_{\shuffle}\,\sum_{\shuffle'}\,C^{\epsilon}_{\rm YS}(\pmb\a_{m-2}\shuffle\pmb\b_{n-m})\,C^{\W\epsilon}_{\rm YS}(\pmb\a'_{m-2}\shuffle'\pmb\b'_{n-m})\nn
& &~~~~~~~~~{\cal A}^{\rm BAS}(i,\pmb\a_{m-2}\shuffle\pmb\b_{n-m},j|i,\pmb\a'_{m-2}\shuffle'\pmb\b'_{n-m},j)\,,~~\label{expan-GR+S}
\eea
where
\bea
C_{\rm YS}^\epsilon(\pmb\a_{m-2}\shuffle\pmb\b_{n-m})=\sum_{S[\pmb\b_{n-m}]}\,\prod_{k=1}^f\,F_{\pmb r_k}\,.~~\label{c-YMS}
\eea
The expression $C_{\rm YS}^{\W\epsilon}(\pmb\a_{m-2}\shuffle\pmb\b_{n-m})$ can be obtained from $C_{\rm YS}^\epsilon(\pmb\a_{m-2}\shuffle\pmb\b_{n-m})$ by replacing $\epsilon_q$ with $\W\epsilon_q$.
As can be seen, $C_{\rm YS}^\epsilon(\pmb\a_{m-2}\shuffle\pmb\b_{n-m})$ is the product of $F_{\pmb r_k}$ with $k\neq0$. When constructing ordered splitting for $\pmb\b_{m-2}$, the first ordered set $\pmb r_1$ must be $\pmb r_1=\{\cdots,b\}$ if the reference ordering is chosen as $\pmb{\cal R}=\{b,\ldots\}$. We emphasize that each $F_{\pmb r_k}$ in \eref{c-YMS} depends on the full merged ordering $\pmb\a_{m-2}\shuffle\pmb\b_{n-m}$, not solely on $\pmb\b_{n-m}$, since momenta $k_{a_q}$ with $a_q\in\pmb\a_{m-2}$ may enters $Z_{r_1^k}$.

Similarly, NLSM amplitudes can be expanded as
\bea
{\cal A}^{\rm NLSM}_n(\pmb\sigma_n)=\sum_{\pmb\a_{n-2}}\,{C}_{\rm N}(\pmb\a_{n-2})\,{\cal A}^{\rm BAS}_n(\pmb\sigma_n|i,\pmb\a_{n-2},j)\,,
~~\label{expan-NLSM}
\eea
where the coefficient $C_{\rm N}(\pmb\a_{n-2})$ is given by
\bea
C_{\rm N}(\pmb\a_{n-2})=\prod_{q=1}^{n-2}\,k_{a_q}\cdot X_{a_q}\,,~~\label{c-NLSM}
\eea
for the given ordered set $\pmb\a_{n-2}=\{a_1,\cdots,a_{n-2}\}$. Here, the combinatorial momentum $X_{a_q}$ denotes the sum for momenta of all external legs to the left of $a_q$ in the ordering $(i,\pmb\a_{n-2},j)$. For example, for the ordering $(1,2,3,4)$, we have $X_2=k_1$, $X_3=k_{12}$.

The expansion \eref{expan-NLSM} can be straightforwardly generalized to the mixed BAS$\oplus$NLSM  case as
\bea
{\cal A}^{{\rm BAS}\oplus{\rm NLSM}}_n(\pmb\sigma_n|i_\phi,\pmb\a_{\phi;m-2},j_\phi;\b_{p;n-m})=\sum_{\pmb\b_{n-m}}\,\sum_{\shuffle}\,{C}_{\rm NS}(\pmb\a_{m-2}\shuffle\pmb\b_{n-m})\,{\cal A}^{\rm BAS}_n(\pmb\sigma_n|i,\pmb\a_{m-2}\shuffle\pmb \b_{n-m},j)\,,
~~\label{expan-NLSM+S}
\eea
where the coefficient is
\bea
{C}_{\rm N}(\pmb\a_{m-2}\shuffle\pmb\b_{n-m})=\prod_{q=1}^{n-m}\,k_{b_q}\cdot X_{b_q}\,,~~~~{\rm for}~\pmb\b_{n-m}=\{b_1,\cdots,b_{n-m}\}\,.
\eea

As proved in \cite{Du:2018khm}, the summation
\bea
\sum_{\pmb\b_{n-m}}\,\sum_{\shuffle}\,\Big(\prod_{q=1}^{n-m}\,k_{b_q}\cdot X_{bq}\Big)\,{\cal A}^{\rm BAS}_n(i,\pmb\a_{m-2}\shuffle\pmb\b_{n-m},j|\pmb\sigma_n)
\eea
vanishes for arbitrary $\pmb\sigma_n$ if $n-m$ is odd. By applying this property to expansion formulas in this section, we immediately see the following well-known and important features: For NLSM amplitudes, the numbers of external particles must be even; For mixed amplitudes BAS$\oplus$NLSM , the numbers of external pions must be even.
\section{$2$-split of pure X amplitudes}
\label{sec-YM-NLSM}

We are now prepared to demonstrate the $2$-split behavior of more general BAS$\oplus$X and pure X amplitudes. The central idea is to express these amplitudes in terms of the special BAS$\oplus$X amplitudes analyzed in Section \ref{sec-BAS+X}, by employing the universal expansions reviewed in Section \ref{sec-expansion}. The desired $2$-split then naturally follow from the $2$-split properties of these BAS$\oplus$X amplitudes. Specifically, we will derive the $2$-split for ordered YM and NLSM amplitudes in Subsections \ref{subsec-YM} and \ref{subsec-NLSM}, respectively, and subsequently establish the $2$-split for unordered GR amplitudes in Subsection \ref{subsec-GR}. 

\subsection{YM amplitudes}
\label{subsec-YM}

The tree-level YM amplitudes ${\cal A}^{\rm YM}_n(i,\pmb A,j,\pmb B\shuffle k)$, which are ordered  compatible with the kinematic condition \eref{kinematic-condi-split-phi3}, exhibit  the following $2$-split,
\bea
{\cal A}^{\rm YM}_n(i,\pmb A,j,\pmb B\shuffle k)\,&\xrightarrow[]{\eref{kinematic-condi-split-YM}\,,\,\eref{kinematic-condi-YM-split1}}&\,\epsilon_k\cdot{\cal J}_{n_1}^{\rm YM}(i,\pmb A,j,\kappa)\,\times \,{\cal J}_{n+3-n_1}^{{\rm Tr}(\phi^3)\oplus{\rm YM}}(j_\phi,\pmb B_g\shuffle\kappa^*_\phi,i_\phi)\,,~~\label{zero-YM}
\eea
under the kinematic constraints:
\bea
\{\epsilon_a,\,k_a\}\,\cdot\,\{\epsilon_b,\,k_b\}=0\,,&~~~~&{\rm with}~a\in A\,,~b\in B\,,~~\label{kinematic-condi-split-YM}
\eea
and
\bea
\epsilon_c\cdot\,\{\epsilon_b,\,k_b\}=0\,,&~~~~&c\in\{i,j,k\}\,.~~\label{kinematic-condi-YM-split1}
\eea
Upon modifying the condition \eref{kinematic-condi-YM-split1} to
\bea
\epsilon_c\cdot\,\{\epsilon_a,\,k_a\}=0\,,&~~~~&c\in\{i,j,k\}\,,~~\label{kinematic-condi-YM-split2}
\eea
the $2$-split \eref{zero-YM} transitions into the alternative form:
\bea
{\cal A}^{\rm YM}_n(i,\pmb A,j,\pmb B\shuffle k)\,&\xrightarrow[]{\eref{kinematic-condi-split-YM}\,,\,\eref{kinematic-condi-YM-split2}}&\,{\cal J}_{n_1}^{{\rm Tr}(\phi^3)\oplus{\rm YM}}(i_\phi,\pmb A_g,j_\phi,\kappa_\phi)\,\times\, \epsilon_k\cdot{\cal J}_{n+3-n_1}^{\rm YM}(j,\pmb B\shuffle\kappa^*,i)\,.~~\label{split-YM}
\eea
In this subsection, we derive the $2$-split relations  \eref{zero-YM} and \eref{split-YM} by employing the expansion structures given in \eref{expan-YM} and \eref{expan-YM+S}.

Specifically, we take the legs $i$ and $j$ in \eref{expan-YM} to be identical to those in \eref{zero-YM}. With this choice, the amplitude ${\cal A}^{\rm YM}_n(i,\pmb A,j,\pmb B\shuffle k)$ can be expanded as
\bea
{\cal A}^{\rm YM}_n(i,\pmb A,j,\pmb B\shuffle k)=\sum_{\pmb A'}\,\sum_{\pmb B'}\,\sum_{\shuffle'_1,\,\shuffle'_2}\,C_{\rm Y}^\epsilon(\pmb A'\shuffle'_1{\pmb B'\shuffle'_2 k})\,{\cal A}^{\rm BAS}_n(i,\pmb A,j,\pmb B\shuffle k|i,\pmb A'\shuffle'_1{\pmb B'}\shuffle'_2 k,j)\,,
~~\label{expan-YM-AB}
\eea
Note that the above expansion \eref{expan-YM-AB} is a direct reformulation of \eqref{expan-YM}. It holds universally and is not dependent on any specific kinematic constraint, including \eqref{kinematic-condi-split-YM}, \eqref{kinematic-condi-YM-split1}, or \eqref{kinematic-condi-YM-split2}.

To proceed, we now impose kinematic conditions \eref{kinematic-condi-split-YM} and \eref{kinematic-condi-YM-split1}. Furthermore, we choose the reference ordering as $\pmb{\cal R}=\{k,\ldots\}$, where $k$ is placed at the lowest position. This choice implies that $\pmb r_1=\{\cdots,k\}$ if  $k$ is not already included in $\pmb r_0$. A key observation is that any $\pmb r_k$ in the ordered splitting $\{\pmb r_0,\cdots,\pmb r_f\}$ cannot simultaneously contain elements from both $A\cup k$ and $B$; otherwise, the corresponding factor $F_{\pmb r_k}$, defined in \eref{F0} or \eref{Fl}, vanishes. In particular, the ordered set $\pmb{r}_0$ and the special $\pmb{r}_1=\{\cdots,k\}$, only contain elements in $A\cup \{k\}$.  

As a result, each ordered splitting naturally divides into two parts: the first corresponds to a reduced ordered splitting of the set $\pmb A'\shuffle'_2 k$, and the second to that of $\pmb B'$. Thus, the coefficient
$C_{\rm Y}^\epsilon(\pmb A'\shuffle'_1{\pmb B'}\shuffle'_2k)$ from \eref{c-YM} factorizes as
\bea
C_{\rm Y}^\epsilon(\pmb A'\shuffle'_1{\pmb B'}\shuffle'_2k)=\Big(\sum_{S[\pmb A'\shuffle'_2k]}\,\prod_{\pmb r_k}\,F_{\pmb r_k}\Big)\,\Big(\sum_{S[\pmb B']}\,\prod_{\pmb r_k}\,F_{\pmb r_k}\Big)\,,~~\label{c-YM-AB}
\eea
where the sums $\sum_{S[\pmb A'\shuffle'_2k]}$ and $\sum_{S[\pmb B']}$ run over reduced ordered splittings of $\pmb A'\shuffle'_2k$ and $\pmb B'$, respectively.

It is worth to emphasize that any $F_{\pmb r_k}$ belonging to $S[\pmb A'\shuffle'_2k]$ is independent of $\pmb B'$ and $\shuffle'_1$. This independence arises because  each $\pmb r_k$ contains only elements from $A\cup k$, and any contribution from $k_b$ with $b\in B$ to $Z_{r_1^k}$ in $F_{\pmb r_k}$ is nullified by the kinematic conditions
\eref{kinematic-condi-split-YM} and \eref{kinematic-condi-YM-split1}. Therefore, we can perform the summations over $\pmb B'$ and $\shuffle'_1$ independently, without affecting these
$F_{\pmb r_k}$. In contrast, each
$F_{\pmb r_k}$ associated with $S[\pmb B']$ may depend on $\pmb A'\shuffle'_2k$ and $\shuffle'_1$, as $k_k$ is allowed to enter $Z_{r_1^k}$.

Plugging the observation \eref{c-YM-AB} and expansion of BAS$\oplus$YM amplitudes  \eref{expan-YM+S} into the YM amplitude expansion \eqref{expan-YM-AB}, we obtain
\bea
& & {\cal A}^{\rm YM}_n(i,\pmb A,j,\pmb B\shuffle k)\,\xrightarrow[]{\eref{kinematic-condi-split-phi3}\,,\,\eref{kinematic-condi-split-YM}\,,\,\eref{kinematic-condi-YM-split1}}\nn
& &\sum_{\pmb A'}\,\sum_{\shuffle'_2}\,\Big(\sum_{S[\pmb A'\shuffle'_2k]}\,\prod_{\pmb r_k}\,F_{\pmb r_k}\Big)\,{\cal A}^{{\rm BAS}\oplus{\rm YM}}_n(i_\phi,\pmb A_\phi,j_\phi,\pmb B_g\shuffle k_\phi|i_\phi,\pmb A'_\phi\shuffle'_2 k_\phi,j_\phi;B_g)\,.~~\label{conver-expanYM}
\eea
%
The BAS$\oplus$YM amplitudes in the above matches the structure of \eqref{keysplit-YM-1}, and by applying the corresponding $2$-split, we obtain the result in \eqref{zero-YM}, where the YM current is defined as
\bea
\epsilon_k\cdot{\cal J}_{n_1}^{\rm YM}(i,\pmb A,j,\kappa)=\sum_{\pmb A'}\,\sum_{\shuffle'}\,\Big(\sum_{S[\pmb A'\shuffle'k]}\,\prod_{\pmb r_k}\,F_{\pmb r_k}\Big)\,{\cal J}^{\rm BAS}_{n_1}(i,\pmb{A},j,\kappa|i,\pmb{A}'\shuffle'\kappa,j)\,,~~\label{current1}
\eea
up to an overall minus sign which can be absorbed by redefining one of the resulting currents. The above definition of current in \eref{current1} is justified by the fact that all $F_{\pmb r_k}$ in \eref{current1} involved depend only on the ordering $(i,\pmb{A}'\shuffle'_2\kappa,j)$, and are independent of elements in $B$.

If we instead impose the alternative condition \eref{kinematic-condi-YM-split2}, then $\pmb r_0$ and $\pmb r_1=\{\cdots k\}$ will only involve elements in $B\cup \{k\}$. The factorization behavior of coefficients in \eref{c-YM-AB} is then modified to
\bea
C_{\rm Y}^\epsilon(\pmb A'\shuffle'_1{\pmb B'}\shuffle'_2k)=\Big(\sum_{S[\pmb A']}\,\prod_{\pmb r_k}\,F_{\pmb r_k}\Big)\,\Big(\sum_{S[\pmb B'\shuffle'_2 k]}\,\prod_{\pmb r_k}\,F_{\pmb r_k}\Big)\,.~~\label{c-YM-AB-2}
\eea
Substituting this into the expansion \eqref{expan-YM-AB} and again using \eqref{expan-YM+S}, we obtain
\bea
& & {\cal A}^{\rm YM}_n(i,\pmb A,j,\pmb B\shuffle k)\,\xrightarrow[]{\eref{kinematic-condi-split-phi3}\,,\,\eref{kinematic-condi-split-YM}\,,\,\eref{kinematic-condi-YM-split2}}\nn
& &\sum_{\pmb B'}\,\sum_{\shuffle'_2}\,\Big(\sum_{S[\pmb B'\shuffle'_2k]}\,\prod_{\pmb r_k}\,F_{\pmb r_k}\Big)\,{\cal A}^{{\rm BAS}\oplus{\rm YM}}_n(i_\phi,\pmb A_g,j_\phi,\pmb B_\phi\shuffle k_\phi|i_\phi,\pmb B'_\phi\shuffle'_2 k_\phi,j_\phi;A_g)\,.~~\label{YM-split2-step}
\eea
Finally, inserting the $2$-split \eref{keysplit-YM-2} into \eref{YM-split2-step}, we recover the split form \eqref{split-YM}, with the current given by
where
\bea
\epsilon_k\cdot{\cal J}_{n+3-n_1}^{\rm YM}(j,\pmb B\shuffle\kappa^*,i)=\sum_{\pmb B'}\,\sum_{\shuffle'}\,\Big(\sum_{S[\pmb B'\shuffle'k]}\,\prod_{\pmb r_k}\,F_{\pmb r_k}\Big)\,{\cal J}^{\rm BAS}_{n+3-n_1}(j,\pmb{B}\shuffle \kappa^*,j|i,\pmb{B}'\shuffle'\kappa^*,j)\,.~~\label{current2}
\eea

Before ending this subsection, let us briefly discuss the properties of the resulting currents in the $2$-split. As seen in \eref{current1} and \eref{current2}, the pure YM vector currents inherit the expansion\footnote{In contrast, above derivation could not give the expansion of the resulting BAS$\oplus$YM current. } of corresponding on-shell amplitudes \eref{expan-YM}, where the BAS amplitudes in basis are replaced by the associated BAS currents. Indeed, the coefficients in the expansions \eqref{current1} and \eqref{current2} are computed using exactly the same rules outlined in Section~\ref{sec-expansion} for on-shell amplitudes. This correspondence has a simple explanation: when applying the rules of Section~\ref{sec-expansion} to construct the coefficients in \eqref{current1} and \eqref{current2}, the only difference is that each momentum $k_k$ in  $Z_{r_1^k}$ is replaced by $k_\kappa=k_k+k_B$. However, since any $\pmb r_k$ contains only elements in $A\cup k$, the  $k_B$ component in $k_\kappa$ is effectively annihilated. As a result, $k_\kappa$ is equivalent to $k_k$. Furthermore, by inspecting the expansion formulas \eqref{current1} and \eqref{current2}, along with the choice of reference ordering and the definition of the kinematic factors $F_{{\pmb r}_k}$, one can observe that the contractions $k_i\cdot k_{\kappa}$, $k_j\cdot k_{\kappa}$, $k_i\cdot k_{\kappa^*}$ and $k_j\cdot k_{\kappa^*}$ do not appear in the expressions of any pure YM current. As discussed in \cite{Feng:2025ofq}, this feature plays a crucial role in understanding the structure of $2$-split via the BCFW on-shell recursion relation.

\subsection{NLSM amplitudes}
\label{subsec-NLSM}
We now turn to the analysis of the $2$-split of ordered tree-level NLSM amplitudes. As mentioned at the end of Section \ref{sec-expansion},  any non-vanishing amplitude must involve an even number of external pions.
The $2$-split for such amplitudes can be described as follows
\bea
{\cal A}^{\rm NLSM}_n(i,\pmb A,j,\pmb B\shuffle k)\,\xrightarrow[]{\eref{kinematic-condi-split-phi3}}\,{\cal J}_{n_1}^{\rm NLSM}(i,\pmb A,j,\kappa)\,\times\,{\cal J}_{n+3-n_1}^{{\rm Tr}(\phi^3)\oplus{\rm NLSM}}(j_\phi,\pmb B_p,\shuffle\kappa^*_\phi,i_\phi)\,,~~\label{split-NLSM-1}
\eea
when the set $A$ contains the odd number of elements ($|A|$ odd), and
\bea
{\cal A}^{\rm NLSM}_n(i,\pmb A,j,\pmb B\shuffle k)\,\xrightarrow[]{\eref{kinematic-condi-split-phi3}}\,{\cal J}_{n_1}^{{\rm Tr}(\phi^3)\oplus{\rm NLSM}}(i_\phi,\pmb A_p,j_\phi,\kappa_\phi)\,\times\,{\cal J}_{n+3-n_1}^{\rm NLSM}(j,\pmb B,\shuffle\kappa^*,i)\,,~~\label{split-NLSM-2}
\eea
when $|A|$ is even.

The above $2$-split can be proved using a method analogous to that in  Subsection \ref{subsec-YM}. As in the YM case, we begin by expanding the NLSM amplitude ${\cal A}^{\rm NLSM}_n(i,\pmb A,j,\pmb B\shuffle k)$ as
\bea
{\cal A}^{\rm NLSM}_n(i,\pmb A,j,\pmb B\shuffle k)=\sum_{\pmb A'}\,\sum_{\pmb B'}\,\sum_{\shuffle'_1,\,\shuffle'_2}\,{C}_{\rm N}(\pmb A'\shuffle'_1\pmb B'\shuffle'_2k)\,{\cal A}^{\rm BAS}_n(i,\pmb A,j,\pmb B\shuffle k|i,\pmb A'\shuffle'_1\pmb B'\shuffle'_2k,j)\,,
~~\label{expan-NLSM-AB}
\eea
based on the expansion \eref{expan-NLSM}. Again, this expansion is simply a reformulation of \eref{expan-NLSM},
thus holds universally. Under the locus \eref{kinematic-condi-split-phi3}, the coefficient $C_{\rm N}(\pmb A'\shuffle'_1{\pmb B'}\shuffle'_2k)$ given in \eref{c-NLSM} simplify to
\bea
C_{\rm N}(\pmb A'\shuffle'_1{\pmb B'}\shuffle'_2k)=\Big(\prod_{a_q\in{\pmb A'}}\,k_{a_q}\cdot X^{{\pmb A'}\shuffle'_2k}_{a_q}\Big)\,\Big(\prod_{b_q\in\pmb B'}\,k_{b_q}\cdot X^{\pmb B'\shuffle'_2k}_{b_q}\Big)\,k_k\cdot X_k\,,~~\label{c-NLSM-AB}
\eea
where $X^{{\pmb A'}\shuffle'_2k}$ can be effectively be evaluated in the ordering $(i,{\pmb A'}\shuffle'_2k,j)$ due to the kinematic constraint \eref{kinematic-condi-split-phi3}. A similar interpretation applies to 
$X^{{\pmb B'}\shuffle'_2k}$.

Suppose that $|A|$ is odd and $|B|$ is even. In this case, the effective part of $X_k$ is $X_k^{{\pmb A'}\shuffle'_2k}$. To see this, we decompose $X_k$ as
\bea
X_k=X_k^{{\pmb A'}\shuffle'_2k}+X_k^{{\pmb B'}\shuffle'_2k}-k_i\,.~~\label{decom-Xk}
\eea
Note that the terms $X_{b_q}^{{\pmb B'}\shuffle'_2k}$ and $X_k^{{\pmb B'}\shuffle'_2k}$ are unaltered by the summation over $\pmb A'$ and $\shuffle'_1$ in \eref{expan-NLSM-AB}. Thus, for the $X_k^{{\pmb B'}\shuffle'_2k}$ part in \eref{decom-Xk}, the summation over $\pmb A'$ and $\shuffle'_1$ yields
${\cal A}^{{\rm BAS}\oplus{\rm NLSM}}_n(i_\phi,\pmb A_p,j_\phi,\pmb B_\phi\shuffle k_\phi|i_\phi,\pmb B'_\phi\shuffle'_2k_\phi,j_\phi;A_p)$,
where the expansion \eref{expan-NLSM+S} has been used. Since $|A|$ is odd, the amplitude must vanish.
For $k_i$ term in the decomposition \eref{decom-Xk}, summing over $\pmb A'\shuffle'_1\pmb B'$ in \eref{expan-NLSM-AB} leads to the amplitude ${\cal A}^{{\rm BAS}\oplus{\rm NLSM}}_n(i_\phi,\pmb A_p,j_\phi,\pmb B_p\shuffle k_\phi|i_\phi,k_\phi,j_\phi;A_p\cup B_p)$, which also vanishes due to $|A|+|B|$ being odd.

Thus, we can reorganize the expansion \eref{expan-NLSM-AB} as
\bea
& &{\cal A}^{\rm NLSM}_n(i,\pmb A,j,\pmb B\shuffle k)\xrightarrow[]{\eref{kinematic-condi-split-phi3}}\nn
& &\sum_{\pmb A'}\,\sum_{\pmb B'}\,\sum_{\shuffle'_1,\shuffle'_2}\,\Big(k_k\cdot X_k^{{\pmb A'}\shuffle'_2k}\,\prod_{a_q\in{\pmb A'}}\,k_{a_q}\cdot X^{{\pmb A'}\shuffle'_2k}_{a_q}\Big)\,\Big(\prod_{b_q\in\pmb B'}\,k_{b_q}\cdot X^{\pmb B'\shuffle'_2k}_{b_q}\Big)\,{\cal A}^{\rm BAS}_n(i,\pmb A,j,\pmb B\shuffle k|i,\pmb A'\shuffle'_1\pmb B'\shuffle'_2k,j)\nn
&=&\sum_{\pmb A'}\,\sum_{\shuffle'_2}\,\Big(k_k\cdot X_k^{{\pmb A'}\shuffle'_2k}\,\prod_{a_q\in{\pmb A'}}\,k_{a_q}\cdot X^{{\pmb A'}\shuffle'_2k}_{a_q}\Big)\,{\cal A}^{{\rm BAS}\oplus{\rm NLSM}}_n(i_\phi,\pmb A_\phi,j_\phi,\pmb B_p\shuffle k_\phi|i_\phi,\pmb A'_\phi\shuffle'_2k_\phi,j_\phi;B_p)\,,
~~\label{expan-NLSM-AB2}
\eea
where the final equality uses the expansion \eref{expan-NLSM+S}. 
 Thus, by substituting the $2$-split \eref{keysplit-1}, one obtains the $2$-split \eref{split-NLSM-1}, with the current
\bea
{\cal J}_{n_1}^{\rm NLSM}(i,\pmb A,j,\kappa)=\sum_{\pmb A'}\,\sum_{\shuffle'}\,\Big(k_k\cdot X_k^{{\pmb A'}\shuffle'k}\,\prod_{a_q\in{\pmb A'}}\,k_{a_q}\cdot X^{{\pmb A'}\shuffle'k}_{a_q}\Big)\,{\cal J}_{n_1}^{\rm BAS}(i,\pmb A,j,\kappa|i,\pmb A'\shuffle'\kappa,j)\,.~~\label{current1-NLSM}
\eea

If $|A|$ is even and $|B|$ is odd, a similar manipulation leads to the $2$-split \eref{split-NLSM-2}, with the current
\bea
{\cal J}_{n+3-n_1}^{\rm NLSM}(j,\pmb B\shuffle\kappa^*,i)=\sum_{\pmb B'}\sum_{\shuffle'}\Big(k_k\cdot X_k^{{\pmb B'}\shuffle'k}\prod_{a_q\in{\pmb B'}}k_{b_q}\cdot X^{{\pmb B'}\shuffle'k}_{b_q}\Big)\,{\cal J}_{n+3-n_1}^{\rm BAS}(j,\pmb B\shuffle\kappa^*,i|i,\pmb B'\shuffle'\kappa^*,j)\,.~~\label{current2-NLSM}
\eea

As in the YM case, the pure NLSM currents in \eref{current1-NLSM} and \eref{current2-NLSM} inherit expansion structure of the corresponding on-shell amplitudes. However, the coefficients cannot be directly evaluated using the rules of Section \ref{sec-expansion}, because $X^{\pmb A'\shuffle'k}_k$ or $X^{\pmb B'\shuffle'k}_k$ always contains $k_i$, which leads to non-vanishing $k_B\cdot k_i$ or $k_A\cdot k_i$. As a result, $k_{\kappa}\cdot X^{\pmb A'\shuffle'k}_k$ (or $k_{\kappa^*}\cdot X^{\pmb B'\shuffle'k}_k$) is not equivalent to $k_k\cdot X^{\pmb A'\shuffle'k}_k$ (or $k_k\cdot X^{\pmb B'\shuffle'k}_k$). 

An additional interesting observation is that the current \eref{current1-NLSM} vanishes in the soft limit $k_{a_q}\to 0$, while the current \eref{current2-NLSM} vanishes in the soft limit $k_{b_q}\to 0$, as implied by the structure of the expansion coefficients. This reflects that  these pure NLSM currents satisfy Adler zeros for on-shell amplitudes, except taking $k_i$ or $k_j$ to be soft\footnote{Since $k_\kappa$ and $k_{\kappa^*}$ are not the on-shell massless momenta, they can not be taken to be soft.}.
\subsection{GR amplitudes}
\label{subsec-GR}

Finally, we consider the $2$-split of tree-level GR amplitudes. The $2$-split of these amplitudes reads
\bea
A^{\rm GR}_n&\xrightarrow[]{\eref{kinematic-condi-split-GR}\,,\,\eref{kinematic-condi-GR-split1}}&\epsilon_k\cdot{\cal J}^{\rm GR}_{n_1}(\{i,j,k\}\cup A)\cdot\W\epsilon_k\,\times\,{\cal J}^{{\rm Tr}(\phi^3)\oplus{\rm GR}}_{n+3-n_1}(i_\phi,j_\phi,\kappa^*_\phi;B_h)\,,~~\label{split-GR}
\eea
where the constraints on kinematic variables are given by
\bea
\{\epsilon_a,\,\W\epsilon_a,\,k_a\}\,\cdot\,\{\epsilon_b,\,\W\epsilon_b,\,k_b\}=0\,,&~~~~&{\rm with}~a\in A\,,~b\in B\,,~~\label{kinematic-condi-split-GR}
\eea
and
\bea
\{\epsilon_c,\,\W\epsilon_c\}\,\cdot\,\{\epsilon_b,\,\W\epsilon_b,\,k_b\}=0\,,&~~~~&c\in\{i,j,k\}\,.~~\label{kinematic-condi-GR-split1}
\eea
It is unnecessary to consider the dual scenario where the indices $b$ in \eqref{kinematic-condi-GR-split1} are replaced by $a$, due to the  symmetry.

The proof of $2$-split in \eref{split-GR} closely parallels the arguments presented earlier. One starts with the expansion \eqref{expan-GR}, expressing GR amplitudes as
\bea
A^{\rm GR}_n=\sum_{\pmb A,\,\pmb A'}\,\sum_{\pmb B,\,\pmb B'}\,\sum_{\shuffle_1,\,\shuffle_2}\,\sum_{\shuffle'_1,\,\shuffle'_2}& &{C}^\epsilon_{\rm Y}(\pmb A\shuffle_1{\pmb B}\shuffle_2 k)\,{C}^{\W\epsilon}_{\rm Y}(\pmb A'\shuffle'_1{\pmb B'}\shuffle'_2 k)\nn
& &{\cal A}^{\rm BAS}_n(i,\pmb A\shuffle_1{\pmb B}\shuffle_2 k,j|i,\pmb A'\shuffle'_1{\pmb B}'\shuffle'_2 k,j)\,.
~~\label{expan-GR-AB}
\eea
Using the same reasoning as in the discussion below Eq.\eqref{c-YM-AB}, one can perform the summations over
$\pmb B$, $\pmb B'$, $\shuffle_1$ and $\shuffle'_1$, which leads to 
\bea
A^{\rm GR}_n\,&\xrightarrow[]{\eref{kinematic-condi-split-phi3}}&\,
\sum_{\pmb A,\,\pmb A'}\,\sum_{\shuffle_2,\,\shuffle'_2}\,\Big(\sum_{S[\pmb A\shuffle_2k]}\,\prod_{\pmb r_k}\,F_{\pmb r_k}\Big)\,\Big(\sum_{S[\pmb A'\shuffle'_2k]}\,\prod_{\pmb r_k}\,F_{\pmb r_k}\Big)\nn
& &~~~~~~~~~~~~~~~~{\cal A}^{{\rm BAS}\oplus{\rm GR}}_n(i_\phi,\pmb A_\phi,\shuffle_2 k_\phi,j_\phi;B_h|i_\phi,\pmb A'_\phi\shuffle'_2k_\phi,j_\phi;B_h)\,.
~~\label{expan-GR-AB2}
\eea
Finally, substituting the $2$-split relation \eqref{keysplit-BAS+GR} into the above expression yields the desired factorization in ~\eqref{split-GR}. Moreover, the expansion of the resulting pure GR current is given by
\bea
\epsilon_k\cdot{\cal J}_{n_1}^{\rm GR}(\{i,j,\kappa\}\cup A)\cdot\W\epsilon_k&=&\sum_{\pmb A,\,\pmb A'}\,\sum_{\shuffle,\,\shuffle'}\,\Big(\sum_{S[\pmb A\shuffle k]}\,\prod_{\pmb r_k}\,F_{\pmb r_k}\Big)\,\Big(\sum_{S[\pmb A'\shuffle' k]}\,\prod_{\pmb r_k}\,F_{\pmb r_k}\Big)\nn
& &~~~~~~~~~~~~~~~~{\cal J}_{n_1}^{\rm BAS}(i,\pmb A\shuffle\kappa,j|i,\pmb A'\shuffle'\kappa,j)\,.~~\label{current1-GR}
\eea
%


\section{Summary}
\label{sec-summary}

In this paper, by extending the diagrammatic method of \cite{Zhou:2024ddy} and leveraging universal expansions, we have reproduced the novel factorization phenomenon known as the $2$-split for tree-level BAS$\oplus$X and pure X amplitudes, where $\mathrm{X}={\mathrm{YM},\mathrm{NLSM},\mathrm{GR}}$. Our derivations provide an intuitive pictorial interpretation of the $2$-split: it can be viewed as the splitting of propagators along the lines $L_{i,v}$ and $L_{j,v}$. For more complicated amplitudes, this mechanism becomes manifest only after expanding them into a suitable basis. Furthermore, we have identified universal expansions for the resulting pure currents arising in these $2$-split, closely analogous to expansions of the corresponding on-shell amplitudes. Although our derivation does not provide explicit expansions for the BAS$\oplus$YM and BAS$\oplus$GR currents, we have verified that they coincide with the corresponding amplitudes. For the BAS$\oplus$NLSM current, its expansion can be understood by the method described in the next paragraph.

We also highlight an alternative approach to accessing $2$-split for a variety of theories excluding GR. Starting from the $2$-split of pure GR amplitudes, all other $2$-split can be generated by applying the transmutation operators introduced in \cite{Cheung:2017ems}, which map GR amplitudes to amplitudes of other theories. As discussed in \cite{Zhou:2024ddy}, under $2$-split kinematics, each such operator factorizes naturally into two components acting separately on the two currents. This procedure thereby produces $2$-split behavior for all tree amplitudes considered in \cite{Cachazo:2014xea,Cheung:2017ems,Zhou:2018wvn}. While this approach depends on having the $2$-split for GR amplitudes as an initial input, the method presented here offers an effective means to obtain this crucial starting point.

As outlined in Section \ref{sec-intro}, a natural and intriguing question is whether the novel $2$-split phenomenon can be extended further—to encompass a broader range of theories at tree level, as well as to loop-level Feynman integrands, potentially with suitable modifications. Addressing this question demands robust methods for deriving $2$-split, and we expect that the approach developed in this paper may serve as a promising candidate.

\section*{Acknowledgments}

This work was supported in part by the Grants No.NSFC-12535003, No.NSFC-12035008 and No.NSFC-12475122, by the Guangdong Major Project of Basic and Applied Basic Research No.2020B0301030008, and by the Fundamental Research Funds for the Central Universities (Grant No. 010-63253121). It was also supported under Grant No.11935013, No.11947301, No.12047502 (Peng Huanwu Center), No.12247103, and No.U2230402. KZ was also supported by the Grant No.NSFC-11805163.


\bibliographystyle{JHEP}

\bibliography{reference}

\end{document}